\documentclass[a4paper,UKenglish,cleveref, autoref, thm-restate]{article}
\usepackage{arxiv}
\usepackage{amsmath, amssymb}
\usepackage{tikz}
\usepackage{tikzscale}
\usepackage{graphicx}
\usetikzlibrary{shapes, arrows, automata, fit}
\usetikzlibrary{positioning}
\usepackage{mathtools}
\usepackage{hyperref}

\usepackage{subfig}
\usepackage{todonotes}
\usepackage{multirow}
\usepackage{tabularx}
\usepackage{xspace}
\newcommand{\tabfig}{.15}

\usepackage{makecell}
\usepackage[linesnumbered, ruled, vlined]{algorithm2e}

\title{A Graph Model and a Layout Algorithm for Knitting Patterns}
\author{ \href{https://orcid.org/0000-0002-0621-5892}{\includegraphics[scale=0.06]{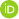}\hspace{1mm}Kathryn Gray} \\
	Department of Computer Science\\
	University of Arizonay\\
	Tucson, AZ 85705 \\
	\url{http://rungray.github.io}\\
	\texttt{ryngray@arizona.edu} \\
	\And
	\href{https://orcid.org/0009-0006-0600-4477}{\includegraphics[scale=0.06]{orcid.pdf}\hspace{1mm}Brian Bell} \\
	Department of Mathematics\\
	University of Arizona\\
	Tucson, AZ 85705 \\
	\texttt{bwbell@math.arizona.edu} \\
	\AND
	Stephen Kobourov \\
        University of Arizona\\
        Tucson, AZ 85705 \\
        \url{https://www2.cs.arizona.edu/~kobourov/}\\
        \texttt{kobourov@arizona.edu} \\
}
\begin{document}

\maketitle

\begin{figure}[ht]
    \centering
    \includegraphics[width=0.95\textwidth]{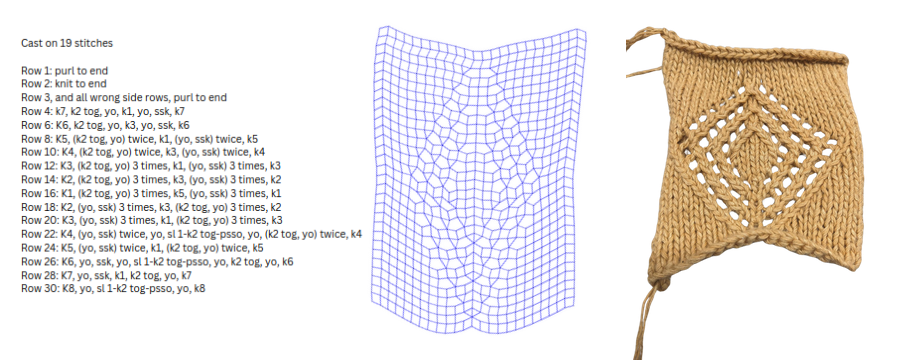}
    \caption{Example of different steps in the process of knitting.  The far left is the written instructions that a knitter would work from.  The center is the output of our algorithm.  The far right is the final, knitted result.}
    \label{fig:teaser}
\end{figure}

\begin{abstract}
 Knitting, an ancient fiber art, creates a structured fabric consisting of loops or stitches. Publishing hand knitting patterns involves lengthy testing periods and numerous knitters. Modeling knitting patterns with graphs can help expedite error detection and  pattern validation. In this paper, we describe how to model simple knitting patterns as planar graphs. We then design, implement, and evaluate a layout algorithm to visualize knitting patterns. Knitting patterns correspond to graphs with pre-specified edge lengths (e.g., uniform lengths, two lengths, etc.). This yields a natural graph layout optimization problem: realize a planar graph with pre-specified edge lengths, while ensuring there are no edge crossings. We quantitatively evaluate our algorithm using real knitting patterns of various sizes against three others; one created for knitting patterns, one that maintains planarity and optimizes edge lengths, and a popular force-directed algorithm. 

\end{abstract}

\section{Introduction}
\label{sec:typesetting-summary}

Knitting is an art whose exact origins are disputed. The earliest probable knitting comes from the 7th to 9th century  \cite{Rutt_history_1987} and represents a simplification from more rudimentary textiles like those produced by naalbinding (labor-intensive threading of fiber with a needle into interlocking loops which do not unravel, unlike knitting which unravels if not tied off) and weaving (which is produced by interleaving fiber in a grid pattern and produces very stiff fabric compared with knitting). Remarkable for the property that knit objects will unravel down to an un-knotted piece of yarn, knit objects are generally equivalent to the un-knot. This simplification makes knit fabrics easier to produce by hand and much more elastic than is possible by other methods. Built on knowledge from the medieval period through the present, it is now a popular hobby. There are generally two types of knitting: warp knitting and weft knitting. Warp knitting is used in machine knitting and consists of several yarns looping around each other. In this paper we focus on weft knitting, also known as hand knitting, which  uses one yarn and consecutively created stitches.

Many knitters rely on knitting patterns: a set of instructions created by the designer of the pattern. An example of a knitting pattern can be seen in Figure~\ref{fig:teaser}. New patterns are created daily, with collections of knitters testing and editing these patterns. Creating and running these tests can be time consuming and difficult. The pattern designer must find people willing to ``test knit'' their proposed pattern by hand and this pool may be small. They must find people who have adequate yarn, time, and knitting background for their pattern. If the pattern is an article of clothing, a wide variety of sizes must be tested. Once the pattern testers have been found, it can be a months-long process for each of the knitters to create the object, while they find and fix any errors. A tool that can take the knitting pattern and output a useful visualization could simplify and expedite this process. Viewing the generated knitting pattern can help catch issues with the knitting such as too many stitches, too few stitches, cables or other adornments in the wrong spot, or creating an unintended shape.

% \begin{figure}

% \centering
% \subfloat[]{\label{fig:knit_pattern}
% \centering
%   \includegraphics[width=0.3\linewidth]{images/knitting_pattern.png}
%     }
% %no space
% \hfill
% \subfloat[]{\label{fig:knit_pattern_knit}
% \centering

% \includegraphics[width=0.3\linewidth]{images/diamond_medalion_knit.jpeg}
% }
% \hfill
% \subfloat[]{\label{fig:knit_pattern_knit}
% \centering

% \includegraphics[width=0.3\linewidth]{images/diamond_medalion.png}
% }

% \caption{\textcolor{black}{A written knitting pattern (a) and the output when the instructions are knit (b), and the output of our algorithm (c).}}
% \end{figure}\label{knit_pattern_example}

In this paper, we begin by showing how to convert a knitting pattern to a graph. We give a complexity definition for different knitting techniques based on how simply the graph can be laid out and how simple the knitting stitch is. For example, many simple knitting patterns correspond to planar graphs, while more complex techniques correspond to non-planar graphs. We will focus on planar knitting and limit our layouts to 2d (flat) representations of knitting which includes a large class of patterns (scarves, shawls, lace fabrics, structured fabrics, etc.). These can be thought of as the locally flat sections of larger garments which may have curvature which is only representable in 3 dimensions.
%-- we reserve these more broad structures for future work. 
Now that we know how to create a knitting graph, what will it look like? Since we focus on the planar graph case, we enforce no edge crossings in the output. We use a force-directed algorithm, as these lead to a natural way to understand the forces the stitches exert on each other. Additionally, we define edge lengths within the pattern which we will attempt to satisfy in order to be consistent with realistic properties of stitches. Knitting stitches are generally slightly taller than wide, and techniques such as drop stitches, short rows, and lace designs can create stitches that are farther or closer than the average stitch. Keeping these close to the definition gives knitting output closer to what the final knitted object will look like.

\section{Previous Work}

We build on work by Counts that models knitting with directed graphs~\cite{counts2018}. This paper generates graphs from knitting patterns and then attempts to visualize the results using a customized heuristic approach, as well as known graph embedding methods such as  multidimensional scaling~\cite{coxon1972multidimensional, torgerson1952multidimensional} and the Kamada-Kawaii algorithm~\cite{kamada1989algorithm}. While Counts  outlines how to get a basic idea of the shape and connectivity of a knitted piece, his heuristic method is not evaluated beyond a high-level, visual comparison.
We consider the theory of knitting graphs, propose an algorithm that takes into account knitting specifics, and propose metrics to evaluate differentalgorithms.

Another way to model knitting is by looking at the individual building blocks in terms of how the yarn is positioned \cite{markande2020}. This paper looks at the topological construction of knitting and breaks it down to the smallest repeatable section. 
While not about knitting, the book ``Figuring Fibers'' contains a chapter~\cite{calderhead_2018} that discusses a method to understand intermeshed crochet using graph theory. Here we see the graphs are used for understanding how the two meshes fit together. A planar graph that can be crochet will give its dual as locations for the intermeshed portion. 

A generative method for creating knitting patterns, KnitGIST\cite{Hofmann_knitgist_2020}, takes different design considerations and stitch patterns to create hand-knitting patterns. This work gives the opposite direction from our paper, starting from a computer image and creating a pattern.

\subsection{Other Fiber Crafts}

Several other fiber crafts have been researched and modeled as graphs including crochet, bobbin lace, and weaving. Each of these share some similarities with knitting, while also differing fundamentally from knitting.

Crochet is the craft that most often gets confused with knittin as the differences can be subtle: crochet can loop the yarn through itself multiple times while knitting loops only once. It keeps only one stitch ``live'' (open) for bringing loops through at a time, while knitting keeps an entire row of stitches live. Crochet has been studied in terms of meshes~\cite{guo_crochetmeshes_2020}. There are also methods for converting a crochet pattern to a graph in order to help the pattern creation process~\cite{seitz_3dcrochet_2021}.

Bobbin lace has been studied in the graph drawning communitiy~\cite{Biedl_bobbin_2017}. It  differs from knitting as it relies on multiple threads that wrap around each other, while knitting is built from one thread that loops around itself.

Weaving has also been modeled by graphs~\cite{akleman_weaving_2009, Hu_weaving_2013}. 
%This thesis looks at weaving by creating a graph to weaving equivalence and defining how to apply a woven fabric to various shapes.  A similar paper uses the shape and yarn structure to create weaving around various shapes\cite{akleman_weaving_2009}.  
Weaving differs from knitting in that weaving usually consists of two sets of threads perpendicular to each other.  These threads go over and under each other to create the woven pattern.

% \subsection{Hand-knitting}

\subsection{Machine Knitting}

Many works look at machine knitting since it is usually easier to make things at scale with this technique. 
One paper that starts to bridge the gap \cite{Hofmann_knitpick_2019} begins at hand knitting and converts to machine knitting. This paper also creates methods to combine stitch techniques while keeping the general shape and stretch of the fabric. 

Another method for generating machine knitting patterns \cite{Liu_4dknitting_2021} looks at specific locations on the garment, such as the shoulders, and where they need to have more or less stretch based on body mechanics. From here, they put different knit stitches in different locations to create fabrics with less strain.

The modeling of knitting using finite automata has also been studied~\cite{grishanov_1997}. This focuses on machine knitting and the state of the needles and yarn in each step. 

Machine knitting has also been modeled using meshes~\cite{narayanan2018}.
This model is similar but not the same as the one we propose in this paper. 
%Several rules for their graphs, several of which we allow in our framework, such as no crossings. Since we are focused on hand-knitting, we have different rules for what consists of feasible knitting. 

Another paper generates machine knitting patterns into arbitrary shapes using the mesh technique \cite{Wu_stitchmeshes_2019}.  This paper \cite{Yuksel_yarnlevel_2012}  creates a yarn-level visualization.  First, it generates a mesh that can correspond to different yarn positions. Then, it relaxes these positions so that the yarn is in a more realistic shape. This technique works for a variety of different stitches.
% We include a more detailed description of their rules and how we modify them in the appendix.

Another visualization technique for creating machine knitting patterns \cite{yu_visualization_2020}, creates a code-based knitting pattern creator paired with a visualization of the outcome. Users are able to find specific stitches and update their code in real time.

There is a paper looking at visually modeling knitting using edge length forces and normal forces to look at different knitting structures \cite{mckinlay_visualization_2023}. This paper begins with a stitch dictionary, defining how a mesh would be created from this. They also give examples of three-dimensional structures. 

\subsection{Planar Graph Visualization}

Realizing planar graphs on a grid without crossings is a well-known problem with a linear-time solution~\cite{chrobak_1995_lineartime}\cite{fraysseix_planar_1990}.  However, such solutions do not take into account pre-specifiededge lengths; see Fig.~\ref{fig:planar_layout}.

PrED~\cite{Bertault_1999} and ImPrED~\cite{Simonetto_2011} are algorithms for force-directed improvements of a given layout.  They improve the given graph layout without changing its topology. In particular, if the input layout is a plane realization of a planar graph, PrED and ImPrED should improe the quality of the layout without introducing edge  crossings. maintain either no-crossings or a given number of crossings.  ImPrED is an update to the PrED algorithm that speeds up the process and allows visualization of larger graphs.

Creating good layouts for trees is related as trees can be realized without edge crossings. Tree layout methods that start with a crossings-free layout and then improve desirable additional criteria (e.g., desired edge length realization) via force-directed methods have been explored in the static case~\cite{Gray_readable_2024} and in the dynamic case~\cite{gray_evolving_2022}. The work in this paper can be seen as generalizing these ideas from trees to planar graphs.

Another approach to edge crossings is to reduce them instead of ensuring that none exist.  It is especially useful when a graph is not guaranteed to be planar. It has been shown that initializing force-directed algorithms and spring embedders with a planar layout can help the final output have fewer edge crossings~\cite{Fowler_preprocessing_2013}. An algorithm that allows both hard and soft constraints, UNICORN~\cite{yu_unicorn_2022}, can lay out a graph with defined types of constraints. 
%These algorithms generally look at reducing edge crossings instead of removing them altogether.  

% \subsection{Visualizing Knitting}

% \subsection{Planar Graph Visualization}

\section{Converting Knitting Patterns to Graphs}
\begin{figure}

\centering
\subfloat[]{\label{fig:single_stitch}
\centering
  \includegraphics[width=0.45\linewidth]{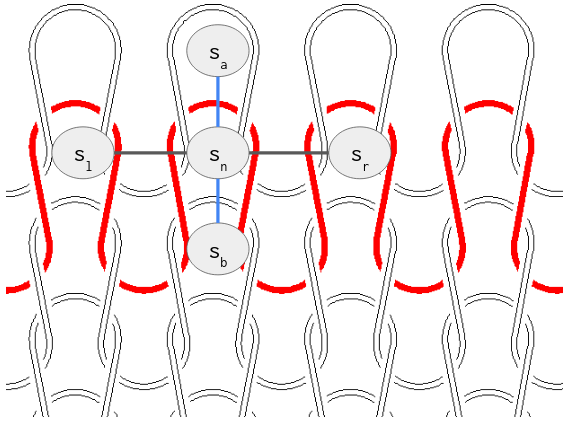}
    }
%no space
\hfill
\subfloat[]{\label{fig:ham_eul}
\centering

\includegraphics[width=0.45\linewidth]{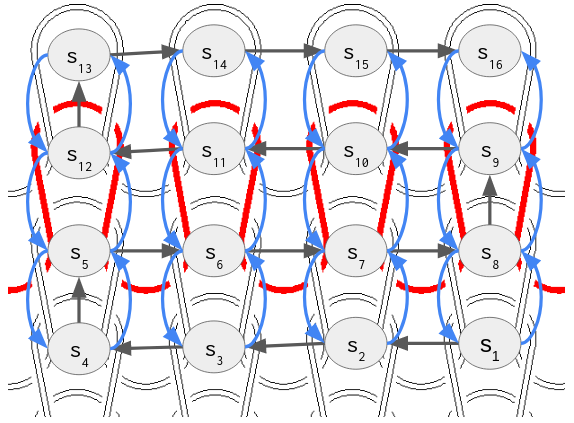}
}

\caption{An example of how we set up the graph.  In (a), we see the graph set up from a single stitch, $S_n$.  $S_n$ connects via yarn with stitches $S_l$ and $S_r$ and connects via loops to $S_a$ and $S_b$.  On the right, in (b), we see an example of the complete graph.  Here, we show the edges along the Hamiltonian path in dark gray, the arrow represents the order of the path.  We show the edges not in the Hamiltonian path but in the Eulerian path in blue.  In both figures, we color one of the rows in red for ease of seeing the row and individual stitches on that row.}
\end{figure}\label{fig:set_up}

 The simplest form of knitting involves a single piece of yarn that is secured at one end, and consists of a series of loops of yarn which are chained through each other (using two needles usually from right to left). 

 % While knitting is generally done with two needles, the gap between the two needles forms a path through every stitch as they are formed as shown in Figure.~\ref{fig:loops}. We will define a graph to represent each knitted object. The \emph{knitting graph} consists of nodes which correspond 1-1 with the loop component of each stitch in the object. Therefore the path of the needle through the work forms a Hamiltonian path through the knitting graph. 
In general, there are two types of connections in knitting. In the first type, we have loops that are connected {\bf along the yarn sequentially}. This can be understood as loops adjacent to each other on the needles. The second type of connection corresponds to loops that are pulled {\bf through each other}, leaving one on the needle, and one hanging on the work. When converting a knitting pattern to a graph, we will focus on the next stitch on the needle. In general, the right needle will be used to pull a (some) new loop(s) of yarn through this stitch which will then prevent this stitch and any dependent on it from unraveling. Then this stitch is removed from the left needle. In out graph representation, the next stitch is $s_b$ and the new loop created is $s_n$. $s_n$ is then connected to the previous stitch $s_r$ (assuming we are knitting right to left on this row) and $s_b$.\ as seen in Fig:~\ref{fig:single_stitch}. As we continue along the knitting, the rest of the connections are added.

Some stitches, such as yarn-overs (yo), make ones (m1L/m1R), or cast-ons (co) do not contain loop connections below them. In this case, we only add yarn connections to the stitches before it. Additionally, some stitches connect to more than one loop below it, such as knit-2-together (k2tog) and other knitting decreases. We create a definition of different knitting stitches including how many stitches they connect below and how many new stitches they create. An example of some stitches from this dictionary are shown in Table~\ref{tab:examples}. This stitch dictionary is used when converting input knitting patters to graphs. The algorithm for converting knitting patters to graphs is summarized in Alg.~\ref{alg:convert-to-grid}.

Knitting stitches are generally taller than they are wide. This gives us two base edge lengths per type of stitch, one that corresponds to following the yarn and one that corresponds to pulling a loop through another. We add these edge lengths as we convert from the knitting pattern to the graph. Additionally some stitches, such as dropped stitches, create larger edge lengths. This affects the yarn connections but does not affect the loop connections.

% \begin{figure}
%     \centering
%     \includegraphics[width=0.6\textwidth,trim={8cm 8cm 8cm 8cm},clip]{images/2.png}
%     \caption{An example of knitting overlaid with its graph. Edges with arrows show the direction of the knitting and create a Hamiltonian Path. Edges without arrows correspond to the looped connections. These two types of edges will have different defined lengths.}
%     \label{fig:knitschematic}
% \end{figure}

\SetAlgoNoLine
\begin{algorithm}[H] 
    \DontPrintSemicolon
    \SetKwFor{For}{for}{do}{end~for}
    \KwIn{Knit pattern (P)}
    \KwOut{Graph (G)}
    G.nodes = None\\
    G.edges = None\\
    n = 0 \tcp{number of nodes}
    needle = Stack\\
    \For{co in P}{ \tcp{Each cast on (co) stitch}
        needle.append(node$_n$)\\
        G.nodes add node$_n$\\
        G.edges add (node$_n$, node$_{n-1}$) \tcp{n $>= 1$}
        n+=1\\
    }
    \For{stitch in P}{
        \For{i in stitch.add}{ \tcp{Each loop to be added}
            G.nodes add node$_n$
            G.edges add (node$_n$, node$_{n-1}$)
            \For{j in stitch.lower\_connections}{ \tcp{Each connected loop below}
                lower\_stitch = needle.pop()\\
                G.edges add (lower\_stitch, node$_n$)\\
            }
            n+=1\\
        }
    }
    \Return G
    \caption{Convert Knit Pattern to Graph}
    \label{alg:convert-to-grid}
\end{algorithm}

\subsection{Knittable Graphs}

%A natural question that arises then is: how can we go the other direction?  Starting from a graph, can we convert that graph into a knitting pattern?  In this paper, we will not go in depth into these questions, however, we discuss a few key ideas on which graphs we can knit.

When knitting, each stitch is seen by the needles twice, once when it is created and once when it is ``finished'' (when a loop is put through it). Generally, we can follow the creation path and this will create a Hamiltonian path in the underlying graph. The Hamiltonian path can be seen by following the gray arrows in Fig:~\ref{fig:ham_eul}. When we are going from general graphs to knitting patterns, we must have a Hamiltonian path in the graph, or in knitting terms, we must have a connected way to create each stitch. 

The other connections in the graph correspond to how the yarn moves through the graph.  This can be seen in Fig.~\ref{fig:ham_eul} by looking at both the gray and blue edges. If a stitch is connected to another stitch by a loop connection, the edge that is created is on an Eulerian path of the graph, but not on the Hamiltonian. When considering the Eulerian path we must allow for parallel edges and thus non-simple graph representations. 
%we give each section of the loop an edge, that is, half the loop creates an edge from stitch 1 to stitch 2 and the other half creates an edge from stitch 2 to stitch 1. 
The Eulerian path is generated by following the Hamiltonian path until we reach an edge with an out-degree greater than one. From here, we follow each blue edge forward and back until there is only one out-edge along the Hamiltonian path. The interpretation in terms of knitting is very straight-forward: the Eulerian path is simply the path of the yarn through the knitting.

\section{Categorizing Knitting}

Knitting has a rich variety of techniques which have been invented through its long history to add texture, elasticity, coloration, complex lace structures, and other artistic and practical embellishments. Many of these techniques have bearing on our ability to lay out the graphs corresponding to these objects. In our experiments, we focus on knitting patterns which correspond to planar knitting, which represents the simplest and most common form of knitting. We discuss how other techniques can determine the minimum number of crossings and the $k$-planarity of knitting graphs and also describe classes of complexity based on the types of additional information needed for graph drawing. A brief summary can be found in Table~\ref{tab:knitting_complexity}. 

We define the simplest knitting patterns and graphs with our complexity class 0. These correspond to planar graphs. In knitting, this includes techniques such as knit and purl stitches, increases, decreases, and simple lace. In the experimental section of this paper, we focus entirely on this type of knitting. We can see the importance of this class by looking at the popular knitting and crochet website, Ravelry \cite{Ravelry}.  Ravelry contains a pattern database where designers can upload patterns and fiber artists can search through them. Looking through Ravelry's pattern database, out of 760,000 patterns, we found approximately 430,000 which include only stitches that can be represented by planar graphs.

\begin{table*}[t]
    \centering
    \begin{tabular}{|c|c|c|c|}
         \hline Stitch  & In Fabric & Subgraph & Edges Added \\ \hline \hline
         \parbox[c]{\tabfig\textwidth}{Knit (k)} &  
         \parbox[c]{\tabfig\textwidth}{\includegraphics[width=\tabfig\textwidth]{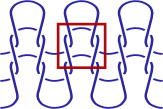}} & 
         \parbox[c]{\tabfig\textwidth}{
\resizebox{\tabfig\textwidth}{!}{         
\begin{tikzpicture}

\node[circle, minimum size=19pt, fill=black!25, inner sep=0pt] (a11) at (0.0, 0.0) {$1$};
\node[circle, minimum size=19pt, fill=black!25, inner sep=0pt] (a21) at (1.3, 0.0) {$2$};
\node[circle, minimum size=19pt, fill=black!25, inner sep=0pt] (a31) at (2.6, 0.0) {$3$};

\node[circle, minimum size=19pt, fill=black!25, inner sep=0pt] (a12) at (0.0, 1.0) {$6$};
\node[circle, minimum size=19pt, fill=red!50, inner sep=0pt] (a22) at (1.3, 1.0) {$5$};
\node[circle, minimum size=19pt, fill=black!25, inner sep=0pt] (a32) at (2.6, 1.0) {$4$};

\node[circle, minimum size=19pt, fill=black!25, inner sep=0pt] (a13) at (0.0, 2.0) {$7$};
\node[circle, minimum size=19pt, fill=black!25, inner sep=0pt] (a23) at (1.3, 2.0) {$8$};
\node[circle, minimum size=19pt, fill=black!25, inner sep=0pt] (a33) at (2.6, 2.0) {$9$};

\draw[line width=0.02cm] (a11) -- (a21) -- (a31) -- 
                         (a32) -- (a22) -- (a12) -- 
                         (a13) -- (a23) -- (a33) ;
\draw[line width=0.02cm] (a32) -- (a33) -- (a23) --
                         (a22) -- (a21) -- (a11) -- (a12);
\draw[line width=0.03cm, color=red!75] (a21) -- (a22) -- (a32);                         

\end{tikzpicture}}} & 
          \parbox[c]{\tabfig\textwidth}{(4, 5)\\ (2, 5)} \\ \hline
         
          \parbox[c]{\tabfig\textwidth}{Purl (p)} & 
         \parbox[c]{\tabfig\textwidth}{\includegraphics[width=\tabfig\textwidth]{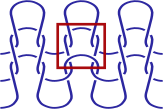}} & 
         \parbox[c]{\tabfig\textwidth}{
\resizebox{\tabfig\textwidth}{!}{         
\begin{tikzpicture}

\node[circle, minimum size=19pt, fill=black!25, inner sep=0pt] (a11) at (0.0, 0.0) {$1$};
\node[circle, minimum size=19pt, fill=black!25, inner sep=0pt] (a21) at (1.3, 0.0) {$2$};
\node[circle, minimum size=19pt, fill=black!25, inner sep=0pt] (a31) at (2.6, 0.0) {$3$};

\node[circle, minimum size=19pt, fill=black!25, inner sep=0pt] (a12) at (0.0, 1.0) {$6$};
\node[circle, minimum size=19pt, fill=red!50, inner sep=0pt] (a22) at (1.3, 1.0) {$5$};
\node[circle, minimum size=19pt, fill=black!25, inner sep=0pt] (a32) at (2.6, 1.0) {$4$};

\node[circle, minimum size=19pt, fill=black!25, inner sep=0pt] (a13) at (0.0, 2.0) {$7$};
\node[circle, minimum size=19pt, fill=black!25, inner sep=0pt] (a23) at (1.3, 2.0) {$8$};
\node[circle, minimum size=19pt, fill=black!25, inner sep=0pt] (a33) at (2.6, 2.0) {$9$};

\draw[line width=0.02cm] (a11) -- (a21) -- (a31) -- 
                         (a32) -- (a22) -- (a12) -- 
                         (a13) -- (a23) -- (a33) ;
\draw[line width=0.02cm] (a32) -- (a33) -- (a23) --
                         (a22) -- (a21) -- (a11) -- (a12);
\draw[line width=0.03cm, color=red!75] (a21) -- (a22) -- (a32);                         

\end{tikzpicture}}} & 
          \parbox[c]{\tabfig\textwidth}{(4, 5)\\ (2, 5)} \\ \hline
          
          \parbox[c]{\tabfig\textwidth}{Yarn over (yo)} & 
          \parbox[c]{\tabfig\textwidth}{\includegraphics[width=\tabfig\textwidth]{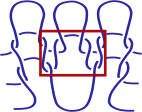}} & 
         \parbox[c]{\tabfig\textwidth}{
\resizebox{\tabfig\textwidth}{!}{         
\begin{tikzpicture}

\node[circle, minimum size=19pt, fill=black!25, inner sep=0pt] (a11) at (0.65, 0.0) {$1$};
\node[circle, minimum size=19pt, fill=black!25, inner sep=0pt] (a21) at (1.95, 0.0) {$2$};

\node[circle, minimum size=19pt, fill=black!25, inner sep=0pt] (a12) at (0.0, 1.0) {$5$};
\node[circle, minimum size=19pt, fill=red!50, inner sep=0pt] (a22) at (1.3, 1.0) {$4$};
\node[circle, minimum size=19pt, fill=black!25, inner sep=0pt] (a32) at (2.6, 1.0) {$3$};

\node[circle, minimum size=19pt, fill=black!25, inner sep=0pt] (a13) at (0.0, 2.0) {$6$};
\node[circle, minimum size=19pt, fill=black!25, inner sep=0pt] (a23) at (1.3, 2.0) {$7$};
\node[circle, minimum size=19pt, fill=black!25, inner sep=0pt] (a33) at (2.6, 2.0) {$8$};

\draw[line width=0.02cm] (a11) -- (a21) -- 
                         (a32) -- (a22) -- (a12) -- 
                         (a13) -- (a23) -- (a33) ;
\draw[line width=0.02cm] (a32) -- (a33) -- (a23) ;
\draw[line width=0.02cm] (a11) -- (a12);
\draw[line width=0.02cm] (a23) -- (a22);

\draw[line width=0.03cm, color=red!75] (a22) -- (a32);                         

\end{tikzpicture}}} & 
          \parbox[c]{\tabfig\textwidth}{(3,4)} \\ \hline
          
          \parbox[c]{\tabfig\textwidth}{Knit Front and Back (kfb)} & 
          \parbox[c]{\tabfig\textwidth}{\includegraphics[width=\tabfig\textwidth]{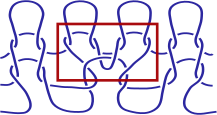}} & 
          \parbox[c]{\tabfig\textwidth}{\resizebox{\tabfig\textwidth}{!}{         
\begin{tikzpicture}

\node[circle, minimum size=19pt, fill=black!25, inner sep=0pt] (a11) at (0.0, 0.0) {$1$};
\node[circle, minimum size=19pt, fill=black!25, inner sep=0pt] (a21) at (1, 0.0) {$2$};
\node[circle, minimum size=19pt, fill=black!25, inner sep=0pt] (a31) at (2, 0.0) {$3$};

\node[circle, minimum size=19pt, fill=black!25, inner sep=0pt] (a12) at (-0.6, 1.0) {$7$};
\node[circle, minimum size=19pt, fill=red!50, inner sep=0pt] (a22) at (0.6, 1.0) {$6$};
\node[circle, minimum size=19pt, fill=red!50, inner sep=0pt] (a32) at (1.6, 1.0) {$5$};
\node[circle, minimum size=19pt, fill=black!25, inner sep=0pt] (a42) at (2.6, 1.0) {$4$};

\node[circle, minimum size=19pt, fill=black!25, inner sep=0pt] (a13) at (-0.6, 2.0) {$8$};
\node[circle, minimum size=19pt, fill=black!25, inner sep=0pt] (a23) at (0.6, 2.0) {$9$};
\node[circle, minimum size=19pt, fill=black!25, inner sep=0pt] (a33) at (1.6, 2.0) {$10$};
\node[circle, minimum size=19pt, fill=black!25, inner sep=0pt] (a43) at (2.6, 2.0) {$11$};

\draw[line width=0.02cm] (a11) -- (a21) -- (a31) -- (a42) -- (a43) -- (a33) -- (a32);
\draw[line width=0.02cm] (a33) -- (a23) -- (a22);
\draw[line width=0.02cm] (a23) -- (a13) -- (a12) -- (a22);
\draw[line width=0.02cm] (a12) -- (a11);
\draw[line width=0.03cm, color=red!75] (a42) -- (a32) -- (a22) -- (a21) -- (a32);    
\end{tikzpicture}}} & 
          \parbox[c]{\tabfig\textwidth}{(4,5), (5,6)\\(2,5), (2,6)} \\ \hline
          
          \parbox[c]{\tabfig\textwidth}{Knit Two Together (k2tog)} & 
          \parbox[c]{\tabfig\textwidth}{\includegraphics[width=\tabfig\textwidth]{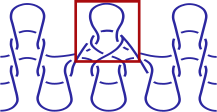}} & 
          \parbox[c]{\tabfig\textwidth}{\resizebox{\tabfig\textwidth}{!}{         
\begin{tikzpicture}

%\node[circle, minimum size=19pt, fill=black!25, inner sep=0pt] (a13) at (0.0, 2.0) {$8$};
%\node[circle, minimum size=19pt, fill=black!25, inner sep=0pt] (a23) at (1, 2.0) {$9$};
%\node[circle, minimum size=19pt, fill=black!25, inner sep=0pt] (a33) at (2, 2.0) {$10$};

\node[circle, minimum size=19pt, fill=black!25, inner sep=0pt] (a12) at (0.0, 1.0) {$9$};
\node[circle, minimum size=19pt, fill=red!50, inner sep=0pt] (a22) at (1, 1.0) {$10$};
\node[circle, minimum size=19pt, fill=black!25, inner sep=0pt] (a32) at (2, 1.0) {$11$};

\node[circle, minimum size=19pt, fill=black!25, inner sep=0pt] (a11) at (-0.6, 0.0) {$8$};
\node[circle, minimum size=19pt, fill=black!25, inner sep=0pt] (a21) at (0.6, 0.0) {$7$};
\node[circle, minimum size=19pt, fill=black!25, inner sep=0pt] (a31) at (1.6, 0.0) {$6$};
\node[circle, minimum size=19pt, fill=black!25, inner sep=0pt] (a41) at (2.6, 0.0) {$5$};

\node[circle, minimum size=19pt, fill=black!25, inner sep=0pt] (a10) at (-0.6, -1.0) {$1$};
\node[circle, minimum size=19pt, fill=black!25, inner sep=0pt] (a20) at (0.6, -1.0) {$2$};
\node[circle, minimum size=19pt, fill=black!25, inner sep=0pt] (a30) at (1.6, -1.0) {$3$};
\node[circle, minimum size=19pt, fill=black!25, inner sep=0pt] (a40) at (2.6, -1.0) {$4$};

\draw[line width=0.02cm] (a11) -- (a21) -- (a31) -- (a41) -- (a32);
%\draw[line width=0.02cm] (a33) -- (a23) -- (a22);
\draw[line width=0.02cm] (a12) -- (a22) -- (a32);
\draw[line width=0.02cm] (a12) -- (a11);
\draw[line width=0.03cm, color=red!75] (a12) -- (a22) -- (a21);  
\draw[line width=0.03cm, color=red!75] (a22) -- (a31);
\draw[line width=0.03cm] (a10) -- (a20) -- (a30) -- (a40);
\draw[line width=0.03cm] (a10) -- (a11);
\draw[line width=0.03cm] (a20) -- (a21);
\draw[line width=0.03cm] (a30) -- (a31);
\draw[line width=0.03cm] (a40) -- (a41);

\end{tikzpicture}}} & 
          \parbox[c]{\tabfig\textwidth}{(6, 10)\\(7, 10), (9, 10)} \\ \hline
           
          \parbox[l]{\tabfig\textwidth}{slip one, knit two together pass slipped stitch over (sl1-k2-psso)} & 
          \parbox[c]{\tabfig\textwidth}{\includegraphics[width=\tabfig\textwidth]{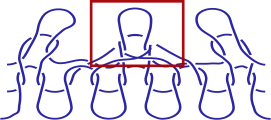}} & 
          \parbox[c]{\tabfig\textwidth}{\resizebox{\tabfig\textwidth}{!}{         
\begin{tikzpicture}

%\node[circle, minimum size=19pt, fill=black!25, inner sep=0pt] (a13) at (0.0, 2.0) {$8$};
%\node[circle, minimum size=19pt, fill=black!25, inner sep=0pt] (a23) at (1.5, 2.0) {$9$};
%\node[circle, minimum size=19pt, fill=black!25, inner sep=0pt] (a33) at (3, 2.0) {$10$};

\node[circle, minimum size=19pt, fill=black!25, inner sep=0pt] (a12) at (0.0, 1.0) {$10$};
\node[circle, minimum size=19pt, fill=red!50, inner sep=0pt] (a22) at (1.5, 1.0) {$11$};
\node[circle, minimum size=19pt, fill=black!25, inner sep=0pt] (a32) at (3, 1.0) {$12$};

\node[circle, minimum size=19pt, fill=black!25, inner sep=0pt] (a11) at (-0.5, 0.0) {$9$};
\node[circle, minimum size=19pt, fill=black!25, inner sep=0pt] (a21) at (0.5, 0.0) {$8$};
\node[circle, minimum size=19pt, fill=black!25, inner sep=0pt] (a31) at (1.5, 0.0) {$7$};
\node[circle, minimum size=19pt, fill=black!25, inner sep=0pt] (a41) at (2.5, 0.0) {$6$};
\node[circle, minimum size=19pt, fill=black!25, inner sep=0pt] (a51) at (3.5, 0.0) {$5$};

\node[circle, minimum size=19pt, fill=black!25, inner sep=0pt] (a10) at (-0.5, -1.0) {$1$};
\node[circle, minimum size=19pt, fill=black!25, inner sep=0pt] (a20) at (0.5, -1.0) {$2$};
\node[circle, minimum size=19pt, fill=black!25, inner sep=0pt] (a30) at (1.5, -1.0) {$3$};
\node[circle, minimum size=19pt, fill=black!25, inner sep=0pt] (a40) at (2.5, -1.0) {$4$};
\node[circle, minimum size=19pt, fill=black!25, inner sep=0pt] (a50) at (3.5, -1.0) {$4$};

\draw[line width=0.02cm] (a11) -- (a21) -- (a31) -- (a41) -- (a51) -- (a32);
%\draw[line width=0.02cm] (a33) -- (a23) -- (a22);
\draw[line width=0.02cm] (a12) -- (a22) -- (a32);
\draw[line width=0.02cm] (a12) -- (a11);
\draw[line width=0.03cm, color=red!75] (a12) -- (a22) -- (a21);  
\draw[line width=0.03cm, color=red!75] (a31) -- (a22) -- (a41);
\draw[line width=0.03cm] (a10) -- (a20) -- (a30) -- (a40) -- (a50);
\draw[line width=0.03cm] (a10) -- (a11);
\draw[line width=0.03cm] (a20) -- (a21);
\draw[line width=0.03cm] (a30) -- (a31);
\draw[line width=0.03cm] (a40) -- (a41);
\draw[line width=0.03cm] (a50) -- (a51);
\end{tikzpicture}}} & 
          \parbox[c]{\tabfig\textwidth}{(6, 11)\\ (7, 11), (8, 11), (10, 11)} \\ \hline
           
          \parbox[c]{\tabfig\textwidth}{Cable One Behind (c1b)} & 
          \parbox[c]{\tabfig\textwidth}{\includegraphics[width=\tabfig\textwidth]{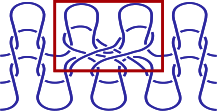}} & 
          \parbox[c]{\tabfig\textwidth}{\resizebox{\tabfig\textwidth}{!}{         
\begin{tikzpicture}

\node[circle, minimum size=19pt, fill=black!25, inner sep=0pt] (a11) at (0.0, 0.0) {$1$};
\node[circle, minimum size=19pt, fill=black!25, inner sep=0pt] (a21) at (1, 0.0) {$2$};
\node[circle, minimum size=19pt, fill=black!25, inner sep=0pt] (a31) at (2, 0.0) {$3$};
\node[circle, minimum size=19pt, fill=black!25, inner sep=0pt] (a41) at (3, 0.0) {$4$};

\node[circle, minimum size=19pt, fill=black!25, inner sep=0pt] (a12) at (0.0, 1.0) {$8$};
\node[circle, minimum size=19pt, fill=black!25, inner sep=0pt] (a22) at (1, 1.0) {$7$};
\node[circle, minimum size=19pt, fill=black!25, inner sep=0pt] (a32) at (2, 1.0) {$6$};
\node[circle, minimum size=19pt, fill=black!25, inner sep=0pt] (a42) at (3, 1.0) {$5$};

\node[circle, minimum size=19pt, fill=black!25, inner sep=0pt] (a13) at (0.0, 2.0) {$9$};
\node[circle, minimum size=19pt, fill=red!50, inner sep=0pt] (a23) at (1, 2.0) {$10$};
\node[circle, minimum size=19pt, fill=red!50, inner sep=0pt] (a33) at (2, 2.0) {$11$};
\node[circle, minimum size=19pt, fill=black!25, inner sep=0pt] (a43) at (3, 2.0) {$12$};

%\node[circle, minimum size=19pt, fill=black!25, inner sep=0pt] (a14) at (0.0, 3.0) {$16$};
%\node[circle, minimum size=19pt, fill=black!25, inner sep=0pt] (a24) at (1, 3.0) {$15$};
%\node[circle, minimum size=19pt, fill=black!25, inner sep=0pt] (a34) at (2, 3.0) {$14$};
%\node[circle, minimum size=19pt, fill=black!25, inner sep=0pt] (a44) at (3, 3.0) {$13$};

\draw[line width=0.02cm] (a12) -- (a11) -- (a21) -- (a31) -- (a41) -- (a42) -- (a32) -- (a31);
\draw[line width=0.02cm] (a42) -- (a43) -- (a33);
%\draw[line width=0.02cm] (a43) -- (a44) -- (a34) -- (a33);
%\draw[line width=0.02cm] (a34) -- (a24) -- (a23);
\draw[line width=0.02cm] (a13) --(a23);
\draw[line width=0.02cm] (a13) -- (a12) -- (a22);
\draw[line width=0.02cm] (a21) -- (a22) -- (a32);
\draw[line width=0.03cm, color=red!75] (a33) -- (a22);
\draw[line width=0.03cm, color=red!75] (a32) --(a23) -- (a33);

\end{tikzpicture}}} & 
          \parbox[c]{\tabfig\textwidth}{(7,11)\\(6,10)} \\ \hline

          \parbox[c]{\tabfig\textwidth}{Cable Two Behind (c2b)} & 
          \parbox[c]{\tabfig\textwidth}{\includegraphics[width=\tabfig\textwidth]{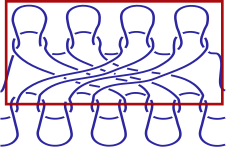}} & 
          \parbox[c]{\tabfig\textwidth}{\resizebox{\tabfig\textwidth}{!}{         
\begin{tikzpicture}

\node[circle, minimum size=19pt, fill=black!25, inner sep=0pt] (a11) at (0.0, 0.0) {$1$};
\node[circle, minimum size=19pt, fill=black!25, inner sep=0pt] (a21) at (1, 0.0) {$2$};
\node[circle, minimum size=19pt, fill=black!25, inner sep=0pt] (a31) at (2, 0.0) {$3$};
\node[circle, minimum size=19pt, fill=black!25, inner sep=0pt] (a41) at (3, 0.0) {$4$};

\node[circle, minimum size=19pt, fill=black!25, inner sep=0pt] (a12) at (0.0, 1.0) {$8$};
\node[circle, minimum size=19pt, fill=black!25, inner sep=0pt] (a22) at (1, 1.0) {$7$};
\node[circle, minimum size=19pt, fill=black!25, inner sep=0pt] (a32) at (2, 1.0) {$6$};
\node[circle, minimum size=19pt, fill=black!25, inner sep=0pt] (a42) at (3, 1.0) {$5$};

\node[circle, minimum size=19pt, fill=red!50, inner sep=0pt] (a13) at (0.0, 2.0) {$9$};
\node[circle, minimum size=19pt, fill=red!50, inner sep=0pt] (a23) at (1, 2.0) {$10$};
\node[circle, minimum size=19pt, fill=red!50, inner sep=0pt] (a33) at (2, 2.0) {$11$};
\node[circle, minimum size=19pt, fill=red!50, inner sep=0pt] (a43) at (3, 2.0) {$12$};

%\node[circle, minimum size=19pt, fill=black!25, inner sep=0pt] (a14) at (0.0, 3.0) {$16$};
%\node[circle, minimum size=19pt, fill=black!25, inner sep=0pt] (a24) at (1, 3.0) {$15$};
%\node[circle, minimum size=19pt, fill=black!25, inner sep=0pt] (a34) at (2, 3.0) {$14$};
%\node[circle, minimum size=19pt, fill=black!25, inner sep=0pt] (a44) at (3, 3.0) {$13$};

\draw[line width=0.02cm] (a12) -- (a11) -- (a21) -- (a31) -- (a41) -- (a42) -- (a32) -- (a31);
\draw[line width=0.02cm] (a12) -- (a22) -- (a21);
\draw[line width=0.03cm, color=red!75] (a43) -- (a33) -- (a12);
\draw[line width=0.02cm] (a22) -- (a32);
%\draw[line width=0.02cm] (a43) -- (a44) -- (a34) -- (a33);
%\draw[line width=0.02cm] (a34) -- (a24) -- (a23);
%\draw[line width=0.02cm] (a24) -- (a14) -- (a13);
\draw[line width=0.03cm, color=red!75] (a33) -- (a23) -- (a42);
\draw[line width=0.03cm, color=red!75] (a23) -- (a13) -- (a32);
\draw[line width=0.03cm, color=red!75] (a43) -- (a22);

\end{tikzpicture}}} & 
          \parbox[c]{\tabfig\textwidth}{(8,11), (7,12)\\(6,9), (5,10)} \\ \hline
    \end{tabular}\medskip
    \caption{Overview of output from all algorithms on general knitting patterns.  The first column defines the knitting stitch we are looking at. The second column shows the actual knitted version.  The boxed area is the stitch and outside is plain knitting. The subgraph column shows the graph of the stitch with some plain stitches around it. We color the stitches of interest and edges created in red to distinguish them. When knitting flat, rows alternate directions, we begin each diagram going left to right. In the last column, we show all the edges that are created in our graph when we encounter the given stitch in a pattern.}
    \label{tab:examples}
\end{table*}

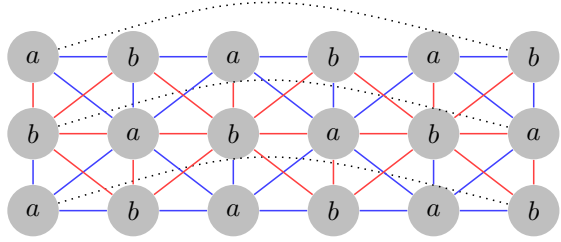
\begin{figure}
    \centering
\resizebox{0.45\textwidth}{!}{         
\begin{tikzpicture}

\node[circle, minimum size=19pt, fill=black!25, inner sep=0pt] (a11) at (0.0, 0.0) {$a$};
\node[circle, minimum size=19pt, fill=black!25, inner sep=0pt] (a21) at (1.3, 0.0) {$b$};
\node[circle, minimum size=19pt, fill=black!25, inner sep=0pt] (a31) at (2.6, 0.0) {$a$};
\node[circle, minimum size=19pt, fill=black!25, inner sep=0pt] (a41) at (3.9, 0.0) {$b$};
\node[circle, minimum size=19pt, fill=black!25, inner sep=0pt] (a51) at (5.2, 0.0) {$a$};
\node[circle, minimum size=19pt, fill=black!25, inner sep=0pt] (a61) at (6.5, 0.0) {$b$};

\node[circle, minimum size=19pt, fill=black!25, inner sep=0pt] (a12) at (0.0, 1.0) {$b$};
\node[circle, minimum size=19pt, fill=black!25, inner sep=0pt] (a22) at (1.3, 1.0) {$a$};
\node[circle, minimum size=19pt, fill=black!25, inner sep=0pt] (a32) at (2.6, 1.0) {$b$};
\node[circle, minimum size=19pt, fill=black!25, inner sep=0pt] (a42) at (3.9, 1.0) {$a$};
\node[circle, minimum size=19pt, fill=black!25, inner sep=0pt] (a52) at (5.2, 1.0) {$b$};
\node[circle, minimum size=19pt, fill=black!25, inner sep=0pt] (a62) at (6.5, 1.0) {$a$};

\node[circle, minimum size=19pt, fill=black!25, inner sep=0pt] (a13) at (0.0, 2.0) {$a$};
\node[circle, minimum size=19pt, fill=black!25, inner sep=0pt] (a23) at (1.3, 2.0) {$b$};
\node[circle, minimum size=19pt, fill=black!25, inner sep=0pt] (a33) at (2.6, 2.0) {$a$};
\node[circle, minimum size=19pt, fill=black!25, inner sep=0pt] (a43) at (3.9, 2.0) {$b$};
\node[circle, minimum size=19pt, fill=black!25, inner sep=0pt] (a53) at (5.2, 2.0) {$a$};
\node[circle, minimum size=19pt, fill=black!25, inner sep=0pt] (a63) at (6.5, 2.0) {$b$};

%\draw[line width=0.02cm] (a11) -- (a21) -- (a31) -- (a41) -- (a51) -- (a61) --
%                         (a62) -- (a52) -- (a42) -- (a32) -- (a22) -- (a12) --
%                         (a13) -- (a23) -- (a33) -- (a43) -- (a53) -- (a63);
\draw[dotted, line width=0.02cm] (a13) .. controls (3, 2.91) .. (a63); 
\draw[dotted, line width=0.02cm] (a12) .. controls (3, 1.9) .. (a62); 
\draw[dotted, line width=0.02cm] (a11) .. controls (3, 0.9) .. (a61);
\draw[line width=0.02cm, color=blue!75] (a13) -- (a23) -- (a33) --
                         (a43) -- (a53) -- (a63) ;
\draw[line width=0.02cm, color=blue!75] (a11) -- (a21) -- (a31) --
                         (a41) -- (a51) -- (a61) ;
\draw[line width=0.02cm, color=red!75] (a12) -- (a22) -- (a32) --
                         (a42) -- (a52) -- (a62) ;                         
\draw[line width=0.02cm, color=red!75]  (a12) -- (a13);
\draw[line width=0.02cm, color=blue!75] (a22) -- (a23); 
\draw[line width=0.02cm, color=red!75]  (a32) -- (a33);
\draw[line width=0.02cm, color=blue!75] (a42) -- (a43);
\draw[line width=0.02cm, color=red!75]  (a52) -- (a53);
\draw[line width=0.02cm, color=blue!75] (a62) -- (a63);

\draw[line width=0.02cm, color=blue!75] (a11) -- (a12);
\draw[line width=0.02cm, color=red!75]  (a21) -- (a22); 
\draw[line width=0.02cm, color=blue!75] (a31) -- (a32);
\draw[line width=0.02cm, color=red!75]  (a41) -- (a42);
\draw[line width=0.02cm, color=blue!75] (a51) -- (a52);
\draw[line width=0.02cm, color=red!75]  (a61) -- (a62);

\draw[line width=0.02cm, color=blue!75] (a11) -- (a22) -- (a31) --
                         (a42) -- (a51) -- (a62) ;                       
\draw[line width=0.02cm, , color=red!75] (a12) -- (a21) -- (a32) --
                         (a41) -- (a52) -- (a61) ; 
\draw[line width=0.02cm, color=blue!75] (a13) -- (a22) -- (a33) --
                         (a42) -- (a53) -- (a62) ;                       
\draw[line width=0.02cm, color=red!75] (a12) -- (a23) -- (a32) --
                         (a43) -- (a52) -- (a63) ;

\end{tikzpicture}}
    \caption{Brioche knitting pattern made by holding 4 strands of yarn in 2 colors. Note that exterior stitches on the sides are connected i.e. ``in the round.''}
    \label{fig:maximal}
\end{figure}

The next complexity class, classs 1, describes knit objects whose graph representation includes crossings and thus their edges must be labeled with an orientation (e.g. front/back). This class of knitting includes cables, color work, drop stitches, and many forms of brioche knitting. We can further refine this class by exploring the $k$-planarity of the underlying knitting graph. We define 1-planar knitting to be exactly all knit objects whose graphs are 1-planar. We can isolate two interesting subclasses based on the types of edges crossed: 1a-planar knitting requires no crossings in edges corresponding with stitches passing through other stitches (``vertical'' edges). 1b-planar knitting requires no crossings in edges corresponding with adjacency between stitches (``horizontal'' edges). Techniques corresponding to 1a-planar knitting include colorwork where yarn color changes every other stitch, drop stitches up to 1 row down, and simple brioche knitting. We note that brioche knitting -- which intentionally introduces extra crossings -- includes at least one technique which generate a maximally 1-planar graph when knitted in the round as shown in Fig.~\ref{fig:maximal}. This technique uses 4 yarns to produce extra crossings. 1b-planar knitting includes single cables (see c1b in Table~\ref{tab:examples} which cross at most 1 column. 

For $k > 2$, we note that the reach of each non-planar technique is increased. Each technique (colorwork, drop stitches, brioche techniques, and cables) has its reach increased by at least $k$ and as much as $2k$, meaning that drop stitches can drop at least $k$ rows down and possibly up to $2k$ rows, depending on the rest of the graph. There is a natural layout of each graph which corresponds to the plane in which it was knit. In this layout, the number of crossings introduced by each technique can be exactly determined by the parameters of the technique. For example in a cable 8 back (c8b), 8 stitches are removed from the needle and held to the back of the knitting while the subsequent 8 stitches are knit and then returned to the needle to be knit. This introduces 8 crossings in the natural knitting layout. However, a different layout may be found which introduces other crossings not naturally in the knitting in order to reduce the number of crossings corresponding with the cable in another layout. 

The complexity class 2 corresponds to knit objects whose vertices also require multiple orientations. This class includes double-knitting, where two layers of fabric (front and back) are knit simultaneously with two yarns allowing the yarns to be swapped arbitrarily between the front and back. This allows for very sophisticated color work and produces a thicker combined fabric. Other techniques,  including some forms of brioche, also fit this classification. Depending on the level of detail required, this class of knitting may require 3d layouts or impose stronger requirements on 2d layouts for visualization. 

%A last consideration is the genus of graphs for each of our classes. Class 0 are planar graphs and require no holes. Class 1 objects are non-planar and thus require at least one hole, thus Class 1 graphs are of genus greater than 1. 1-planar knitting has genus exactly 1 (\cite{Schumacher1981}), however $k$-planar knitting may have genus less than $k$. It may be possible to measure genus by keeping count and orientation information from each non-planar knitting operation. 

\begin{table}[]
    \centering
    \begin{tabular}{|c|c|c|}
        \hline Complexity & Knit Description & Graph Description  \\\hline
        0 &  Common simple knitting & planar graphs \\ \hline
        1 & $k$-planar knitting & \parbox{3.4cm}{Edges have Two Orientations (cables, colorwork, drop-stitches, brioche} \\ \hline
        2 & Double Knitting & \parbox{3.4cm}{Edges and Nodes have\\ multiple orientations}  \\ \hline
        
    \end{tabular}
    \caption{Description of categories of knitting in relation to their graph description.}
    \label{tab:knitting_complexity}
\end{table}

\section{Knitting Graph Layout Algorithm}

In order to layout a graph representation of a knitting pattern, we will perfnorm an initial layout of the graph with no crossings (step 1) and then iteratively improve this layout in order to attempt to satisfy pre-specified edge lengths with a hard constraint against introducing new crossings (step 2). We specifically choose only patterns for which a crossing-free layout is possible and for which a crossing-free initial layout is possible. Note here that knitting patterns contain information that can be used for producing better initial layouts. 
Our goal here is to produce a layout that closely matches knit fabric which has consistent edge lengths and also, due to resistance of yarn to bending, tends to prefer a planar equilibrium shape. The form of our algorithmic approach is summarized in the following pseudocode where ``safe\_FDA\_Step'' is our force directed algorithm step subject to the hard consttraint against edge crossings:

\medskip
\SetAlgoNoLine
\begin{algorithm}[H] 
    \DontPrintSemicolon
    \SetKwFor{For}{for}{do}{end~for}
    \KwIn{Planar Graph (G) with prespecified edge lengths}
    \KwOut{Planar Graph Layout with no crossings}
    $Pos$ = Planar\_Graph\_Layout(G)\tcp{Section 5.1}
    \For{i $<$ iterations}{
        $Pos$ = Safe\_FDA\_Step($Pos$)\tcp{Section 5.2}
    }
    \Return $Pos$
    \caption{KnitLayout}
    \label{alg:alg1}
\end{algorithm}

\subsection{Initial Layout}

\begin{figure}
    \centering
    \includegraphics[width=\columnwidth]{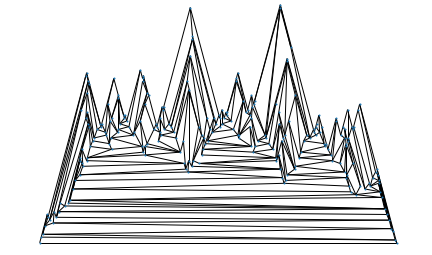}
    \caption{Initialization with a planar layout, this is the pinecone lace knitting pattern.}
    \label{fig:planar_layout}
\end{figure}

For the initial layout, we use the NetworkX~\cite{Hagberg_networkx_2008} planar graph layout. This layout algorithm begins by finding the face with the largest number of nodes and making it the outer face. For our test graphs, this was sufficient to create good layouts as the longest face is usuall the boundary of the knit graph. The NetworkX planar layout is based on the layout algorithm by Chrobak and Payne~\cite{chrobak_1995_lineartime}. which computes positions on a small integer grid; see Fig:~\ref{fig:planar_layout}. We note that this layout does not take advantage of the rich structure of knitting which may allow for more efficient initializations. 

\subsection{Safe Force-Directed Improvement}

The next step is to update this drawing using a force-directed algorithm. We focus on an edge length force. Since we have edge lengths defined for each type of edge, realizing edge lengths while maintaining planarity should result in good representation of the kntitted graph. With this in mind, we aim to realize the desired edge lengths, while enforcing the constraint that no edges cross each other. This hard constraint helps us visualize knitting patterns without folds or out-of-order stitches.

\begin{figure}
    \centering
    \includegraphics[scale=0.3]{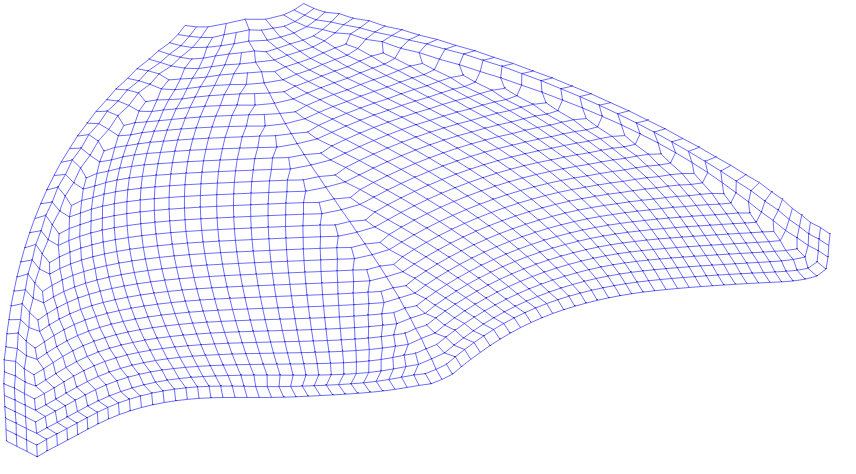}
    \caption{An output from the algorithm, the knitting pattern is a triangle shawl.}
    \label{fig:triangle}
\end{figure}

We ensure that no edge crossings are introduced at any step in the algorithm. First, all new locations are calculated. Then, one by one, nodes are moved to these new positions. If an edge crossing is introduced by this move, that node is kept in place. An output from this can be seen in Fig:~\ref{fig:triangle}.

We include the following forces.  While our main goal is optimizing edge lengths, the other forces help the algorithm converge.

1. The Edge Length Force: this force is attractive if the edge is longer than the defined length and repulsive if it is shorter.

2. Collision Force: we do not want nodes to end up one top of each other. This force keeps nodes far enough apart.

3. Universal repulsive electrostatic force: At the first steps after the initial layout algorithm, nodes tend to move or begin close together. To couunteract this, we apply a universal electrostatic repulsive force between all pairs of vertices in the graph to ensure that nodes move apart. 

\section{Evaluation}
When trying to model knitting graphs, we can define the edge lengths carefully to reflect the edge lengths in knitting. In general, knit stitches are taller than they are wide, which means that the edges that correspond to the rows should be shorter than those corresponding to the columns. 

\begin{figure}
    \centering
    \includegraphics[scale=0.1]{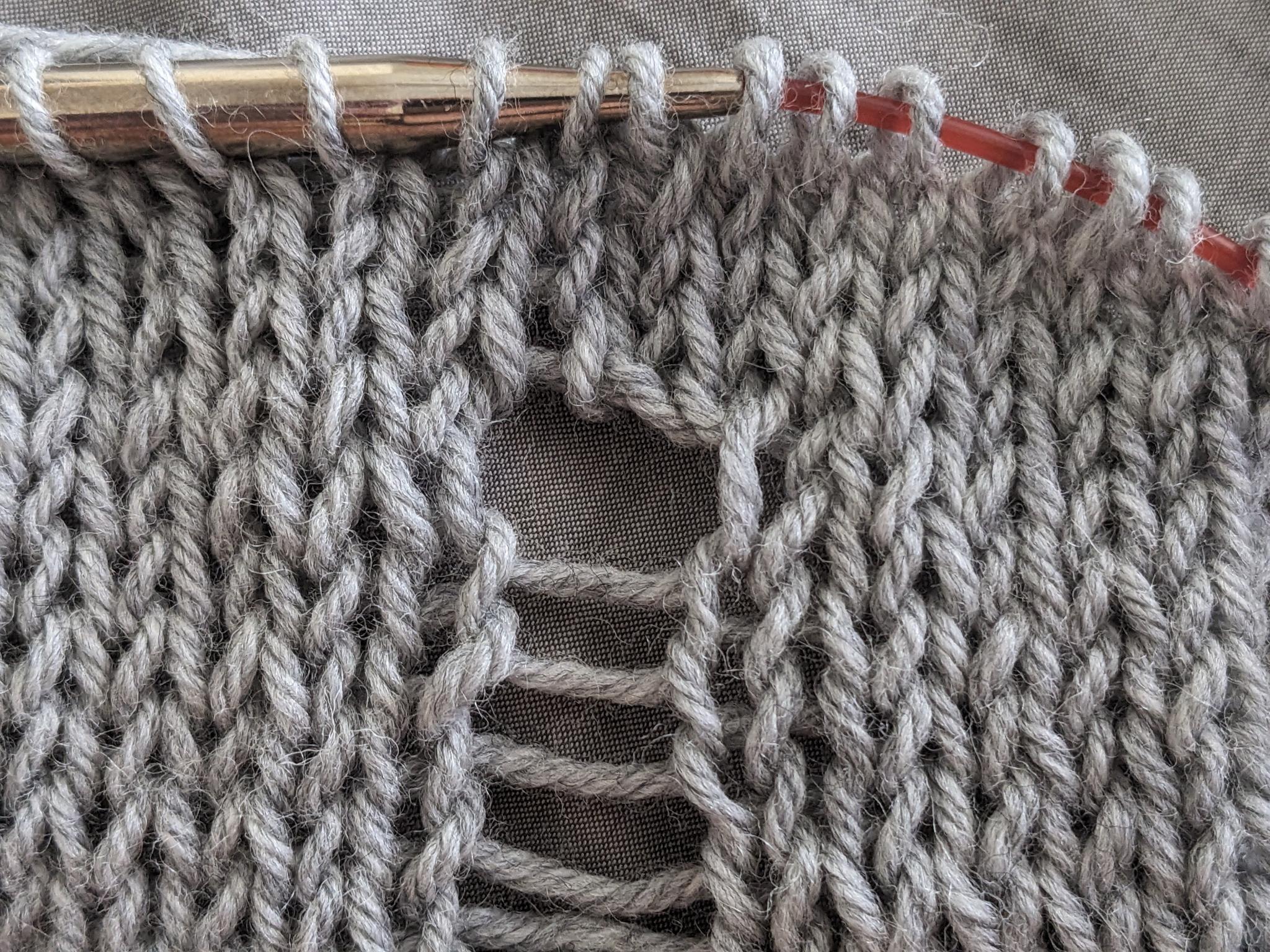}
    \caption{An example of a knitting stitch where edge lengths matter. Here we show a dropped stitch, the edges corresponding to the long, horizontal threads will need to be defined with larger edge lengths.}
    \label{fig:dropped_stitch}
\end{figure}

We can extend this to edges of many types. For example, when two stitches are knit together, one stitch is often longer than the other, causing it to lean in one direction. This fact is often used in lacework to create curved lines. In another example, dropped stitches create a temporary stitch that is removed to create a stitch's length of yarn between two remaining stitches. Fig:~\ref{fig:dropped_stitch} shows a dropped stitch and how the knitting can vary if we change edge lengths.

In our experiments, we include two different knitting graph experiments. One contains a group of many different knitting patterns. The other contains one simple knitting pattern, but the number of rows (and therefore stitches or nodes) is increasing.

First, the group of knitting patterns contains seven different knitting patterns.  Four of these are lace patterns, two are drop stitch patterns, and one is a short row pattern.  The number of nodes ranges from  90 to 1183.  

These knitting patterns come from knittingfool.com~\cite{knittingfool}, some small examples of dropped stitches are created by the authors, and aPincha shawl~\cite{pincha}. 
%Knittingfool.com contains a knitting stitch dictionary that contains about a hundred different knitting stitch patterns. Of these, we focus on the patterns that fall under category 1 of our knitting complexity chart or patterns that correspond to planar graphs.

The second group consists of a triangle that is growing in size.  This is a common, simple pattern that the authors knew from experience.  We begin with five rows and 89 stitches.  Then, we increase the knitting rows by 6 for each new graph.  We end up with a pattern with 1,539 nodes and 35 rows.  This pattern was chosen to showcase a simple common knitting pattern for a triangle shawl.  
%While many of them will have lace or other features, we keep to a basic triangle shawl to show how the algorithms scale in terms of number of nodes.

\subsection{Comparison Algorithms}

We compare the performace of our customized knttings graph layout algorithm against SFDP \cite{Hu_sfdp_2005}, ImPrEd \cite{Simonetto_2011}, and the algorithm by Counts~\cite{counts2018} which we will refer to as ``KnitGrid''.

\begin{figure}
    \centering
    \includegraphics[width=\columnwidth]{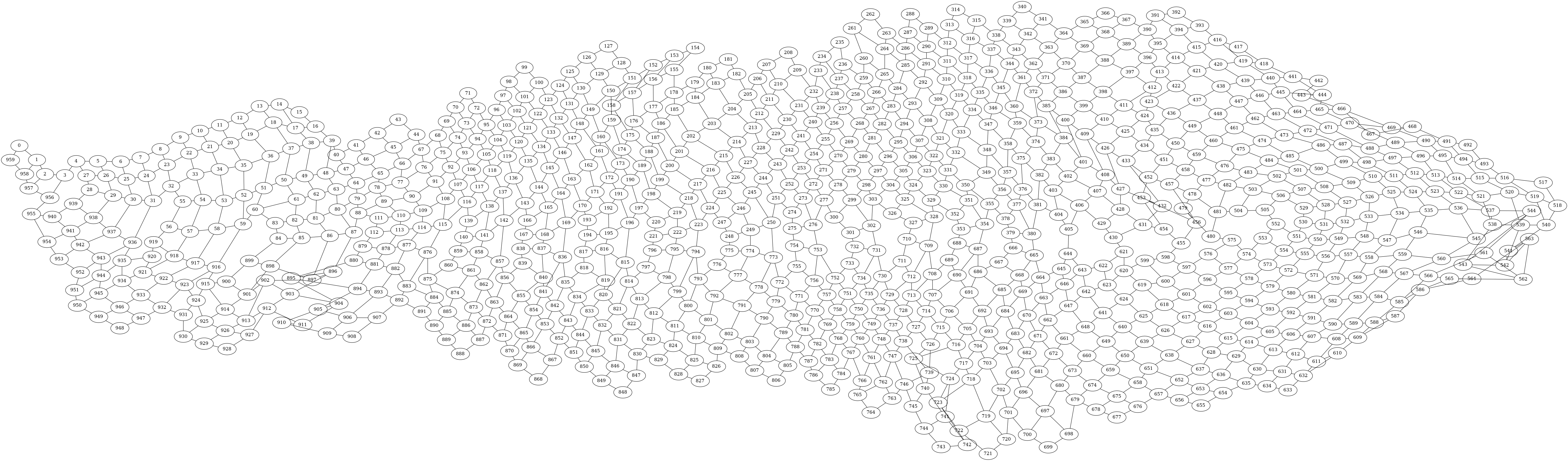}
    \caption{Output of SFDP with edge crossings. Short rows make this graph leaf-shaped.}
    \label{fig:pincha-crossed}
\end{figure}

SFDP is a state-of-the-art scalable force directed placement method. However, it does not attempt to minimize crossings so  it can introduce crossings; see Fig:~\ref{fig:pincha-crossed}.  Nevertheless, as SFDP produces high-quality outputs, especially on smaller graphs, we include this in our comparisons. We use the SFDP implementation in Graphviz~\cite{Gansner_graphviz_2000}, version 2.43.0 together with the prism overlap-removal.

PrEd and ImPrEd improve a given layout without changing its topology. In particular, both PrED and ImPrED prevent edge crossings, if given a crossings-free initial layout. However, these methods are slow and difficult to tune for different graphs. We use the latest implrementation of ImPrEd in Tulip~\cite{Auber_Tulip_2004} version 5.7.4 for these experiments. In our examples, with the same starting graphs, ImPrEd sometimes introduced crossings and got stuck. We include the outputs from ImPrEd and note that this may be an implementation issue in Tulip or ImPrEd (or due to our inability to make them work as intended).

Finally, we compare against the the algorithm of Counts~\cite{counts2018}. This algorithm follows the knitting pattern and places each node on a grid point. It is fast and simple, although it cannot handle all knitting stitches and techniques. For example, it cannot handle short rows, extra cast on or offs, and drop stitches. We use the implementation provided by the author and export node locations for the evaluation.

\subsection{Comparison Techniques}

\textcolor{black}{\textbf{Desired Edge Lengths:}  We compare the different methods  using the well known desired edge length measure (\cite{gray2024, haunert2011drawing, gansner2005graph, dogrusoz2009layout}). 
Given the  desired edge lengths $\{l_{ij}: (i,j) \in E\}$ (here we use edge lengths defined by the knitting), and coordinates of the nodes $X$ in the computed layout, we define the following as the desired edge length measure:}
\begin{align}
    \text{DEL} &= \sqrt{ \frac{1}{|E|} \sum\limits_{(i,j) \in E}\;  
    \left(\frac{||X_i - X_j|| - l_{ij}}{l_{ij}}\right)^2} \label{eq:loss-desired-edge-length}
    \end{align}
    This measures the root mean square of the relative error, % as in~\cite{ahmed2020gd}, 
    producing a positive number, with $0$ corresponding to perfect preservation.

\begin{table}[]
    \centering
    \begin{tabular}{|c|c|c|c|c|}
    \hline
        Pattern & DEL & SFDP & ImPrEd & KnitGrid \\ \hline\hline
         Antique Diamonds & 0.07 & 0.33 & 0.55 & 0.20\\ \hline
         Drop Stitch 1 & 0.42 & 0.68 & 1.04 & 0.64 \\ \hline
         Drop Stitch 2 & 0.44 & 0.67 &  1.1 & 0.53\\ \hline
         Ears of Grass & 0.09 & 0.39 & DNF & 0.18 \\ \hline
         Embossed Double Leaf 1 & 0.15 & 0.40 & DNF & 0.33\\ \hline
         Embossed Double Leaf 2 & 0.20 & 0.24 & DNF & 0.33\\ \hline
         Pincha & 0.31 & 0.39 & DNF & DNC \\ \hline 
         Triangle 1 & 0.06 & 0.27 & 0.9 & 0.64 \\ \hline
         Triangle 2 & 0.15 & 0.36 & 0.62 & 0.51 \\ \hline
         Triangle 3 & 0.21 & 0.34 & DNF & 0.49 \\ \hline
         Triangle 4 & 0.08 & 0.35 & DNF & 0.29\\ \hline 
         Triangle 5 & 0.16 & 0.41 & DNF & 0.28 \\ \hline
    \end{tabular}
    \medskip
    \caption{Edge Length realization comparisons on knitting graphs. We include DNF for ``Did not Finish" for algorithms that were able to start running the graph, but did not finish and DNC for ``Did not Compute" for algorithms that were not able to create the graph.}
    \label{tab:edge_length_knit}
\end{table}
\textcolor{black}{Edge lengths are shown in Table~\ref{tab:edge_length_knit}. Our algorithm outperforms the others by a significant amount. We note that many of these algorithms do not optimize edge lengths.}

\textcolor{black}{\textbf{Runtime:}  We also measure the runtime of each algorithm. Here we show that, while our algorithm runs much slower than SFDP, we can get good output in a reasonable amount of time.}

\begin{table}[]
    \centering
    \begin{tabular}{|c|c|c|c|c|}
    \hline 
        Pattern & DEL & SFDP &  ImPrEd & KnitGrid\\ \hline\hline
         Antique Diamonds & 25 & 0.074  & 0.33 & 0.015\\\hline
         Drop Stitch 1 & 48 & 0.044 & 0.23 & 0.016 \\\hline
         Drop Stitch 2 & 35 & 0.046 & 0.24 & 0.014 \\\hline
         Ears of Grass & 1094 & 0.37 & DNF & 0.095 \\\hline
         Embossed Double Leaf 1 & 1045 & 0.31 & DNF & 0.089\\ \hline
         Embossed Double Leaf 2 & 12303 & 0.66 & DNF & 0.21\\ \hline
         Pincha & 14444 & 0.66 & DNF & DNC\\ \hline
         Triangle 1 & 38 & 0.083 & 8.36 & 0.013 \\ \hline
         Triangle 2 & 118 & 0.15 & 51.5 & 0.031 \\ \hline
         Triangle 3 & 684 & 0.40 & DNF & 0.052 \\ \hline
         Triangle 4 & 3897 & 0.63 & DNF & 0.082 \\\hline
         Triangle 5 & 40906 & 1.01 & DNF & 0.065 \\\hline
    \end{tabular}
    \medskip
    \caption{Comparison of runtimes for the various algorithms in seconds.}
    \label{tab:runtime_knit}
\end{table}

\textcolor{black}{Runtime comparison is shown in Table~\ref{tab:runtime_knit}. While our algorithm is the slowest, it runs in a reasonable amount of time and is able to compute good, crossing-free layouts for all graphs.}

% \begin{table}[]
%     \centering
%     \begin{tabular}{c|c|c|c|c}
%          Graph & node number & Starting Edges & mid& Ending edges  \\
%          Acorn-lace& 154 & 11.0608 &  2.864 & 0.1167
%     \end{tabular}
%     \caption{Caption}
%     \label{tab:my_label}
% \end{table}

\begin{table*}[t]
    \centering
    \begin{tabular}{|c|c|c|c|c|}
         \hline Pattern  & DEL & SFDP & Impred & KnitGrid\\ \hline \hline
         \parbox[c]{\tabfig\textwidth}{Antique\\ Diamonds \\(99 nodes)} &  \parbox[c]{\tabfig\textwidth}{
      \includegraphics[width=\tabfig\textwidth]{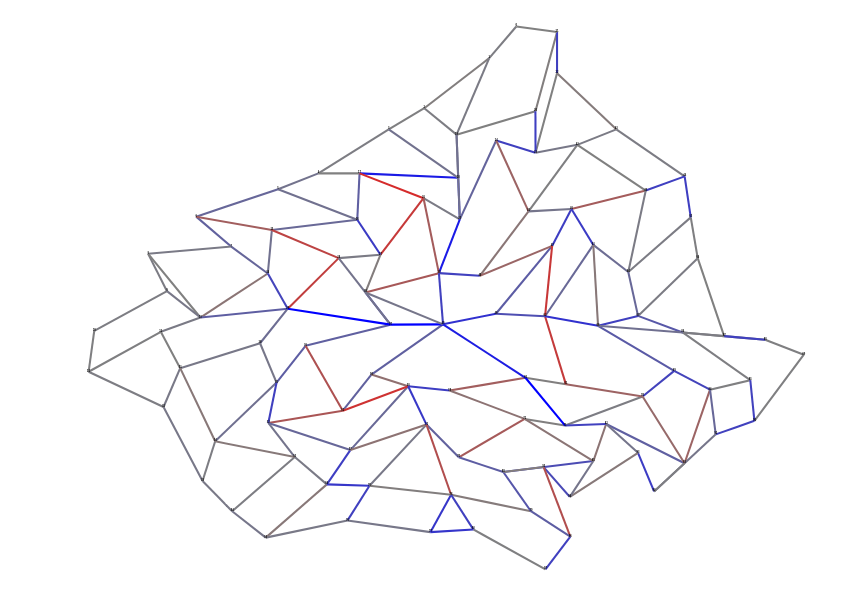}}      & \parbox[c]{\tabfig\textwidth}{
      \includegraphics[width=\tabfig\textwidth]{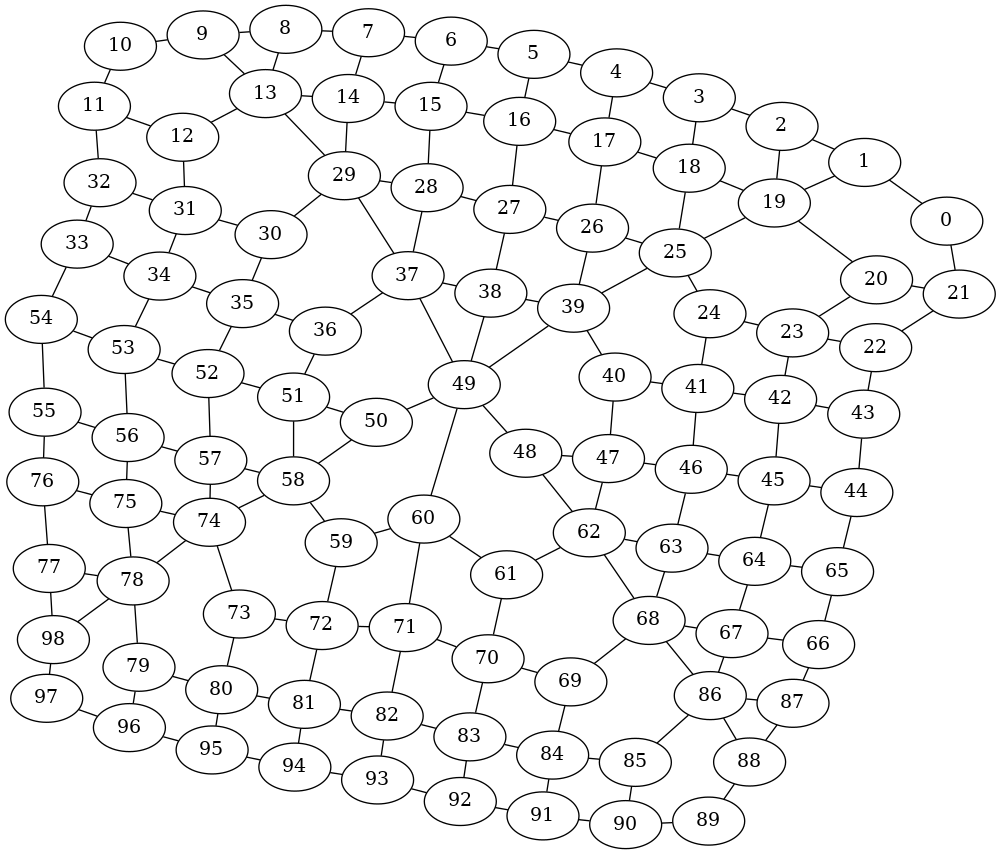}}    
          & \parbox[c]{\tabfig\textwidth}{
      \includegraphics[width=\tabfig\textwidth]{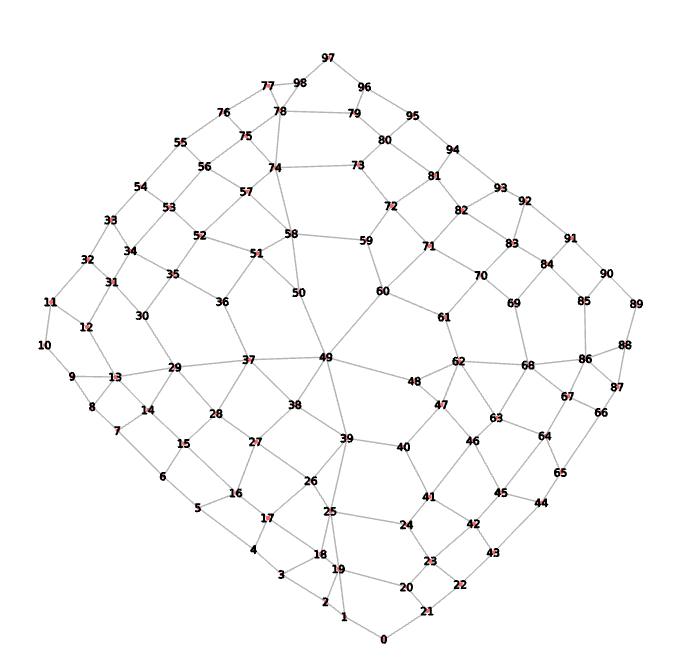}}    
          &  \parbox[c]{\tabfig\textwidth}{
      \includegraphics[width=\tabfig\textwidth]{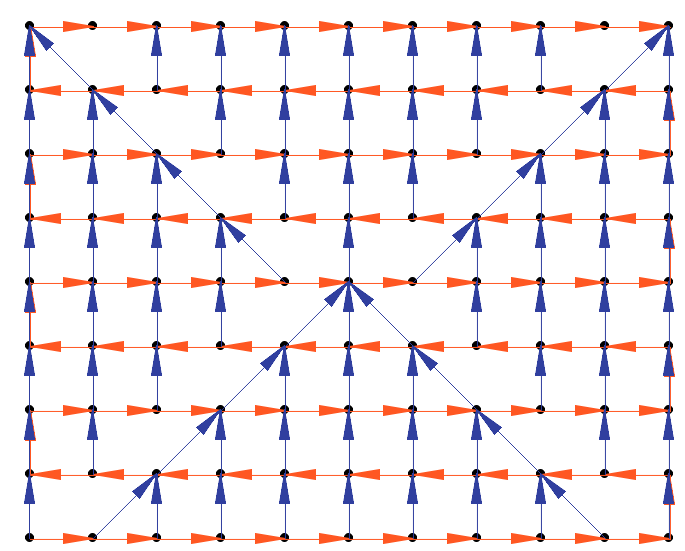}} \\ \hline
          \parbox[c]{\tabfig\textwidth}{Drop Stitch 1 \\ (90 Nodes)} & \parbox[c]{\tabfig\textwidth}{\includegraphics[width=\tabfig\textwidth]{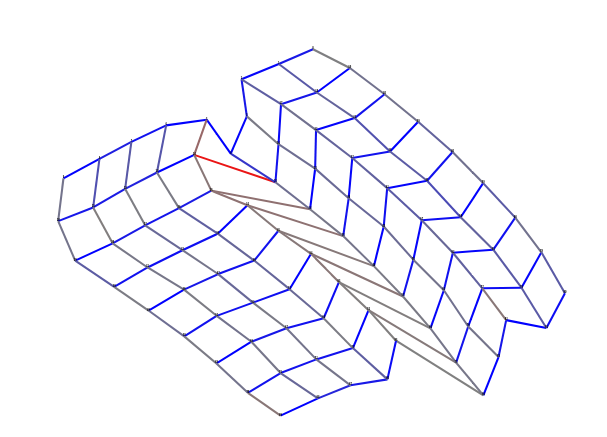}} & 
          \parbox[c]{\tabfig\textwidth}{\includegraphics[width=\tabfig\textwidth]{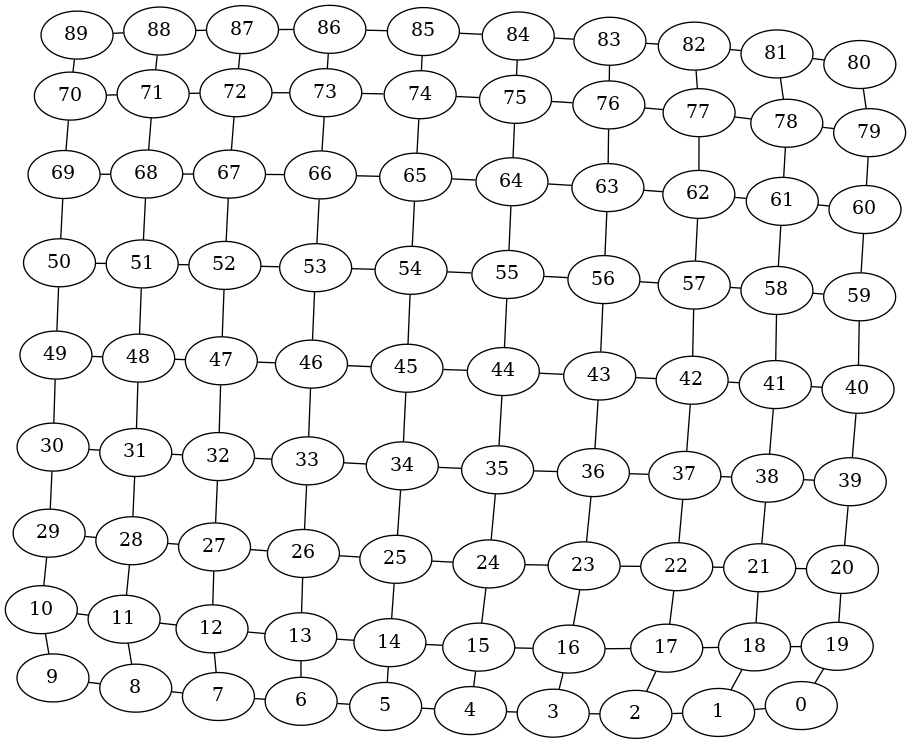}} & \parbox[c]{\tabfig\textwidth}{\includegraphics[width=\tabfig\textwidth]{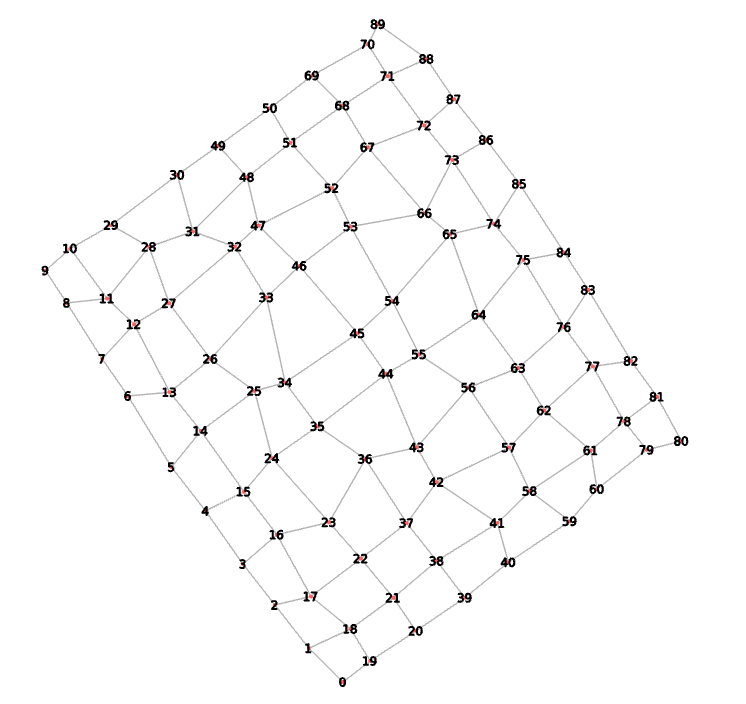}} & \parbox[c]{\tabfig\textwidth}{\includegraphics[width=\tabfig\textwidth]{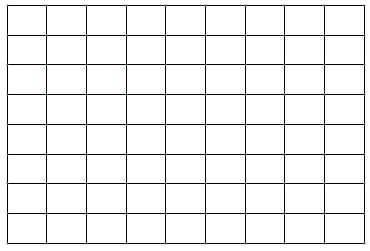}} \\ \hline
          \parbox[c]{\tabfig\textwidth}{Drop Stitch 2 \\ (90 Nodes)} & \parbox[c]{\tabfig\textwidth}{\includegraphics[width=\tabfig\textwidth]{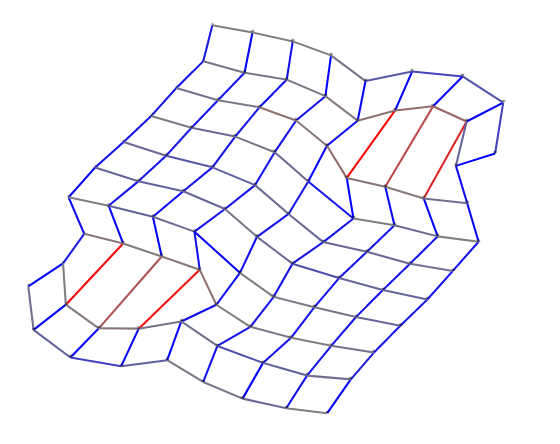}} &
          \parbox[c]{\tabfig\textwidth}{\includegraphics[width=\tabfig\textwidth]{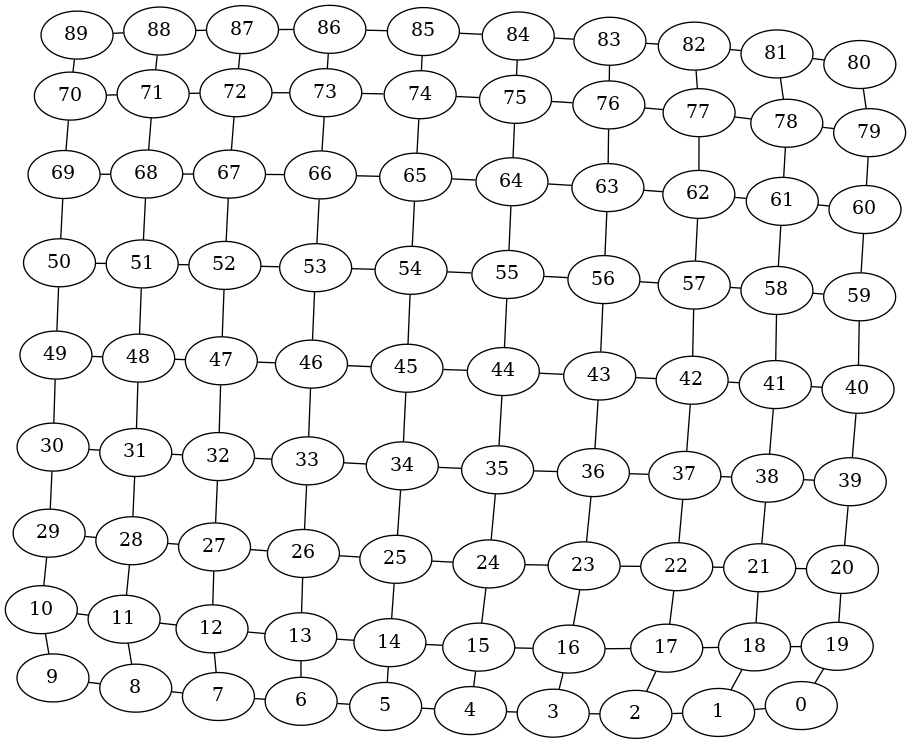}}&\parbox[c]{\tabfig\textwidth}{\includegraphics[width=\tabfig\textwidth]{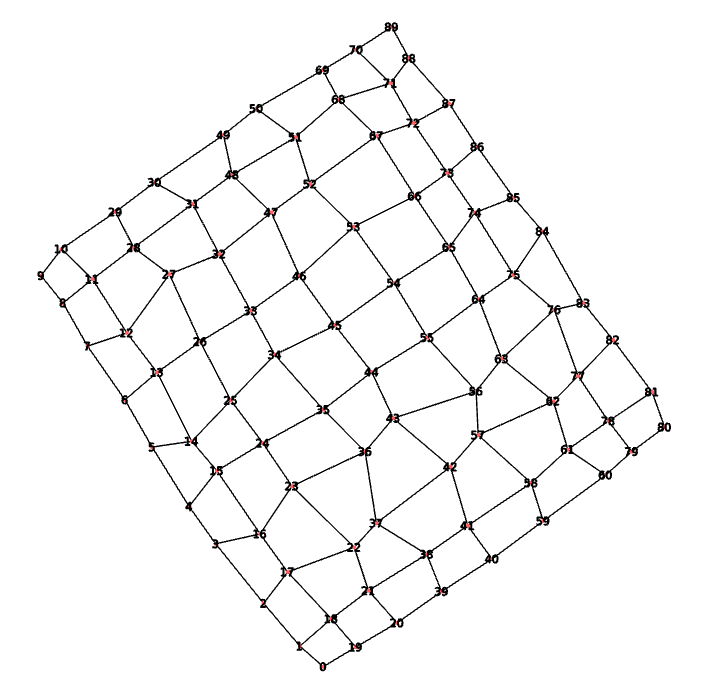}} & \parbox[c]{\tabfig\textwidth}{\includegraphics[width=\tabfig\textwidth]{images/drop_stitch_knitgrid.png}} \\ \hline
          \parbox[c]{\tabfig\textwidth}{Ears of Grass \\(600 nodes)} & \parbox[c]{\tabfig\textwidth}{ \includegraphics[width=\tabfig\textwidth]{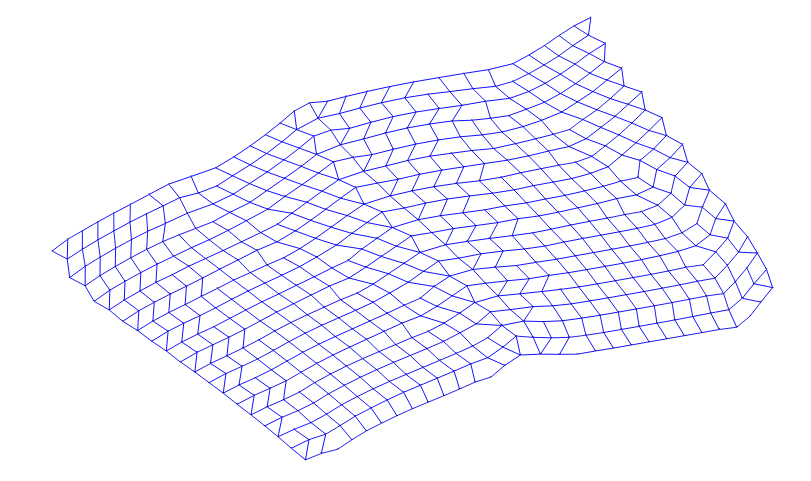}}  &\parbox[c]{\tabfig\textwidth}{\includegraphics[width=\tabfig\textwidth]{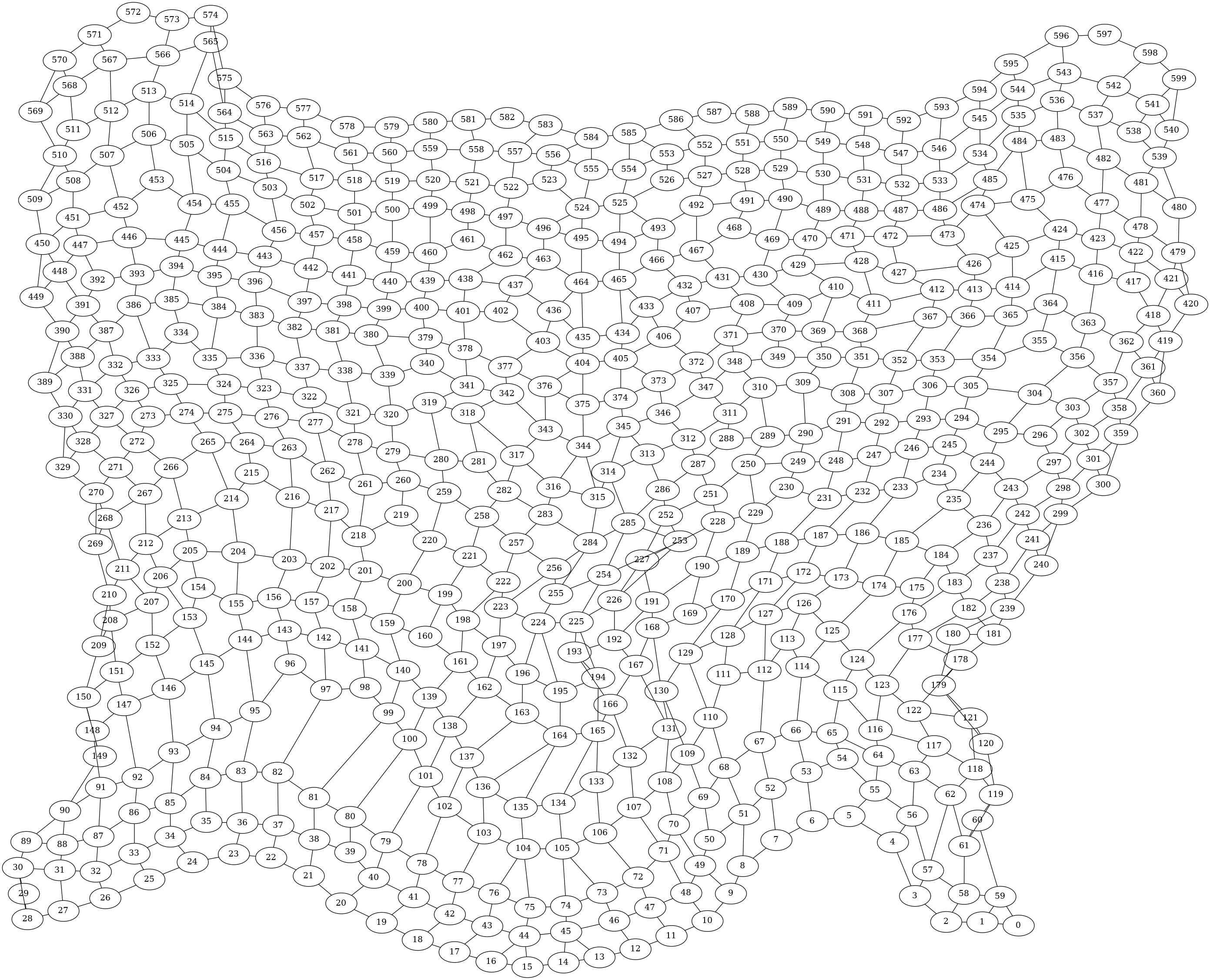}}& \parbox[c]{\tabfig\textwidth}{\includegraphics[width=\tabfig\textwidth]{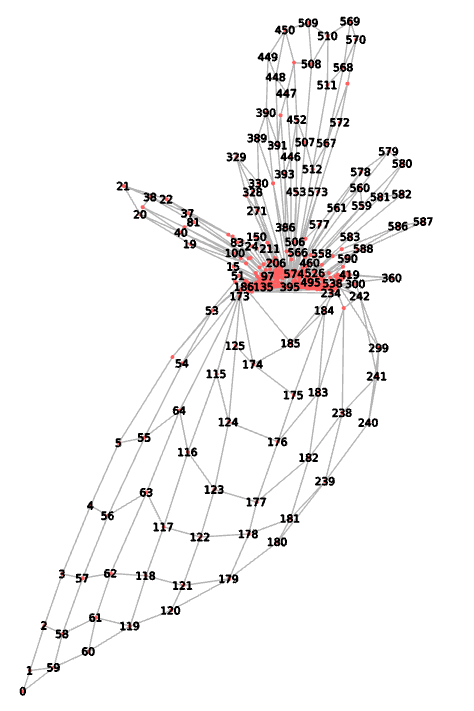}} &\parbox[c]{\tabfig\textwidth}{\includegraphics[width=\tabfig\textwidth]{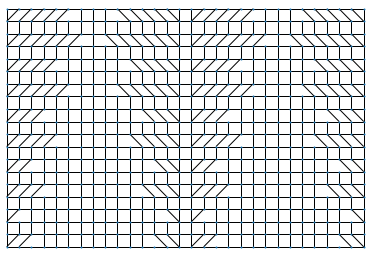}} \\ \hline
          \parbox[c]{\tabfig\textwidth}{Embossed \\Double Leaf \\(606 nodes)} & \parbox[c]{\tabfig\textwidth}{ \includegraphics[width=\tabfig\textwidth]{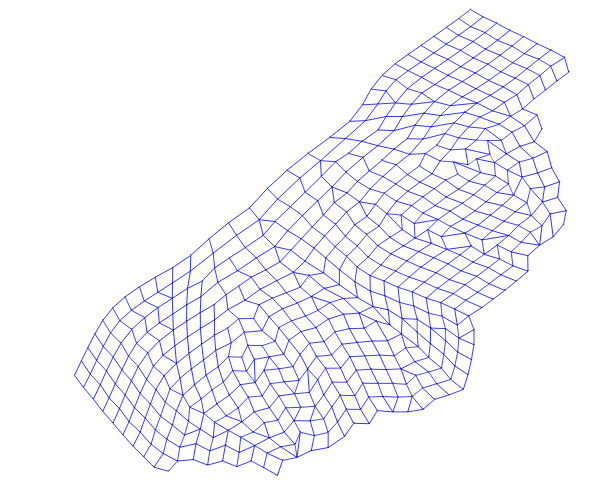}\\}  &\parbox[c]{\tabfig\textwidth}{\includegraphics[width=\tabfig\textwidth]{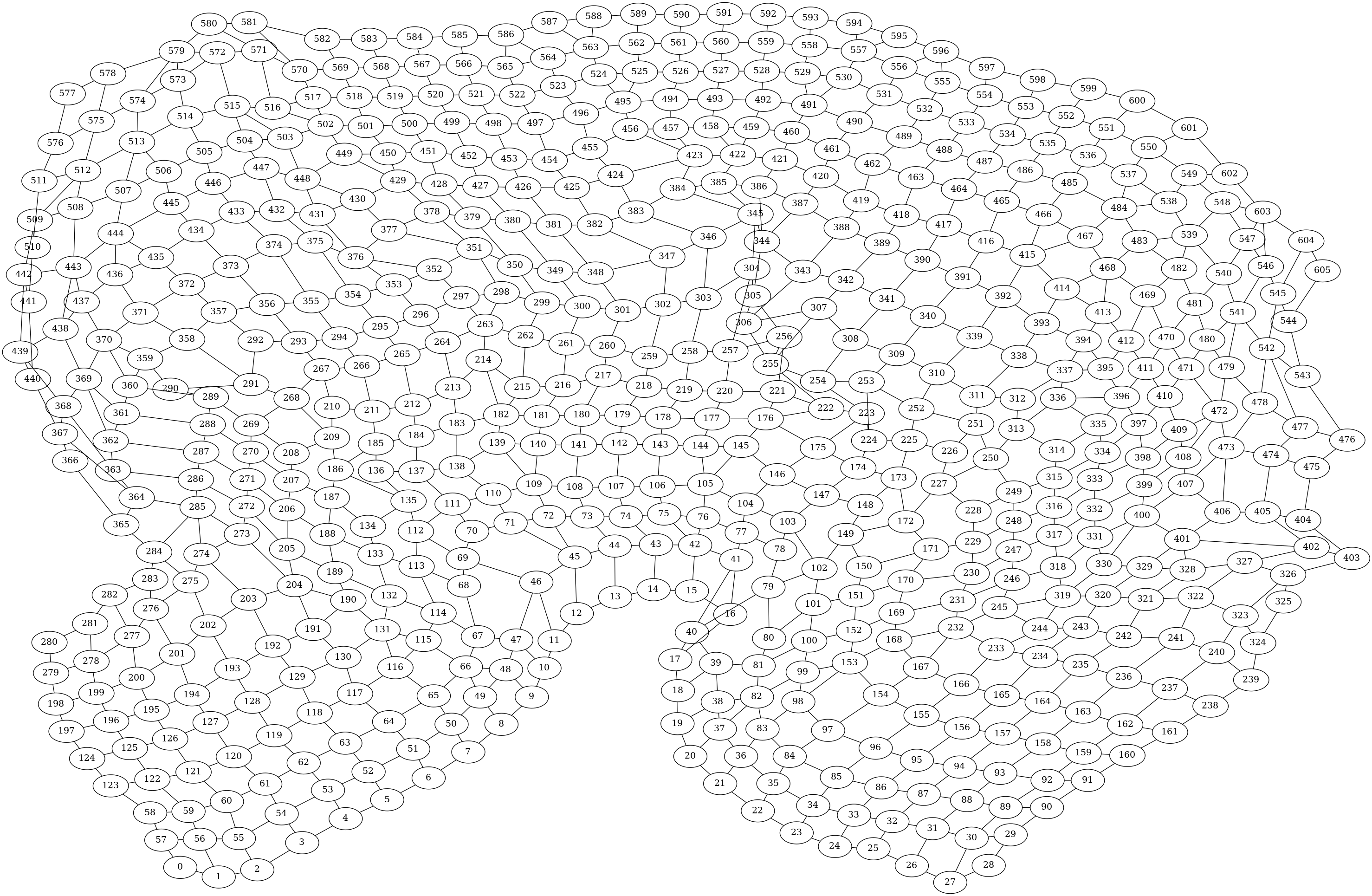}\\ }  
          & DNF
          &\parbox[c]{\tabfig\textwidth}{\includegraphics[width=\tabfig\textwidth]{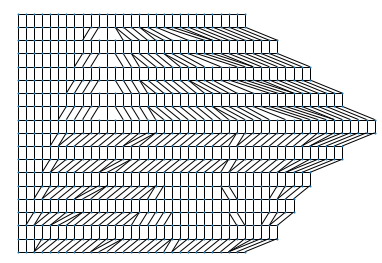}} \\ \hline
          \parbox[c]{\tabfig\textwidth}{Embossed \\ Double Leaf 2 \\(1183 nodes)} & \parbox[c]{\tabfig\textwidth}{ \includegraphics[width=\tabfig\textwidth]{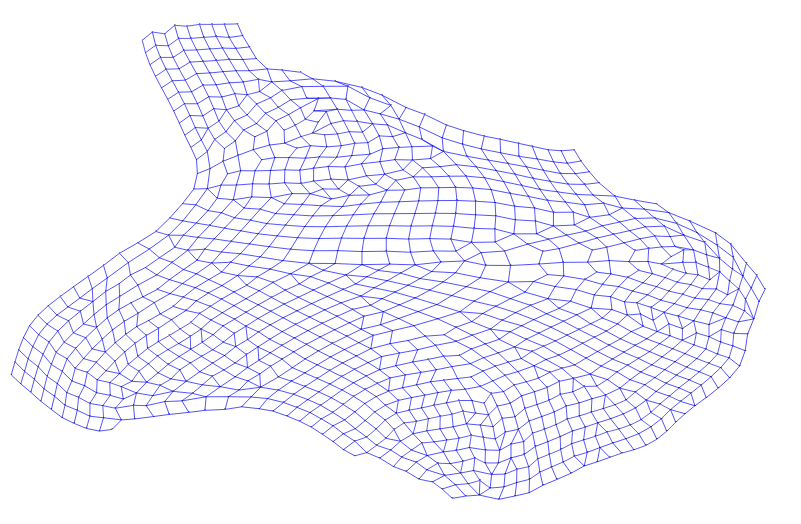}\\}  &\parbox[c]{\tabfig\textwidth}{\includegraphics[width=\tabfig\textwidth]{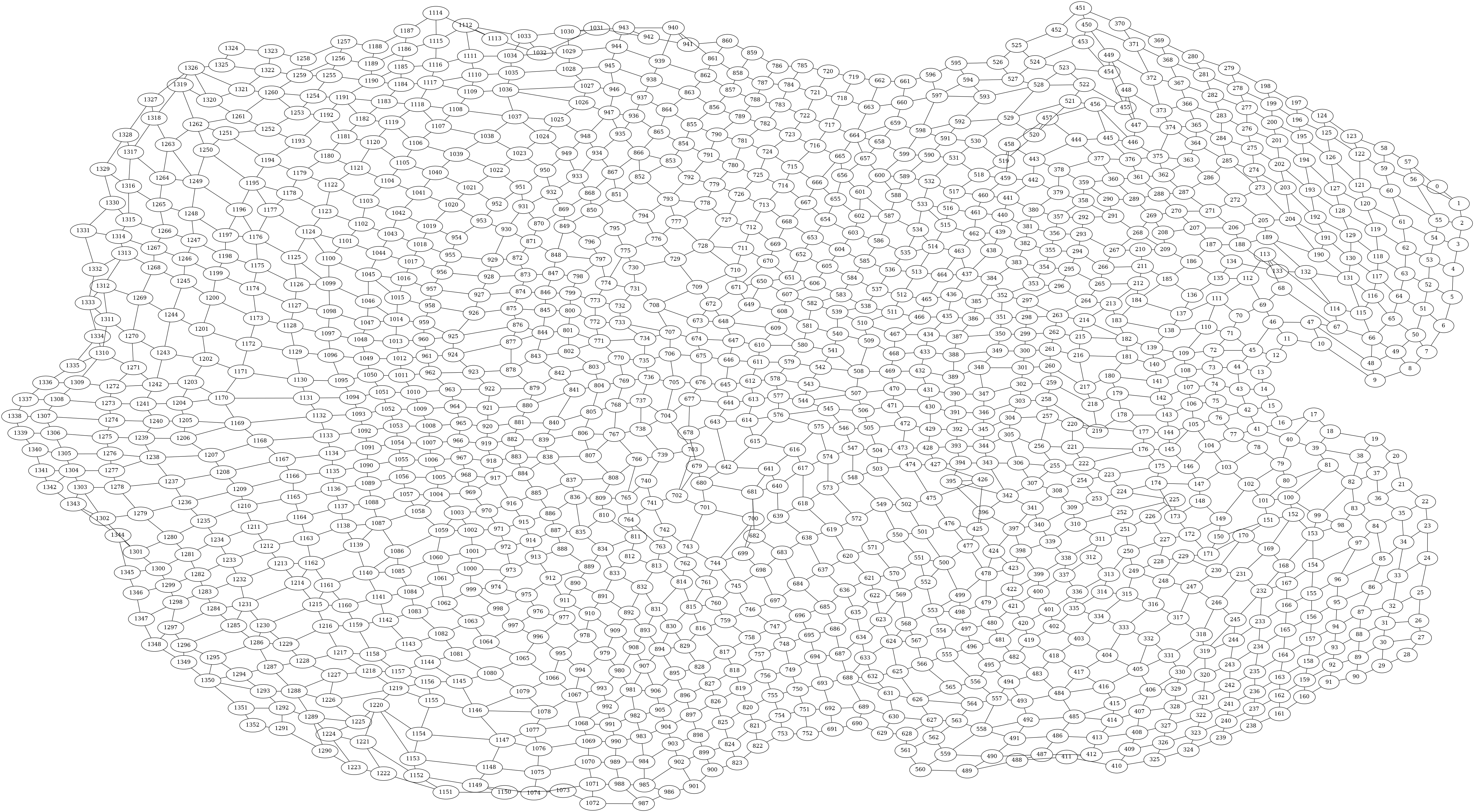}\\ }  
          & DNF
          &\parbox[c]{\tabfig\textwidth}{\includegraphics[width=\tabfig\textwidth]{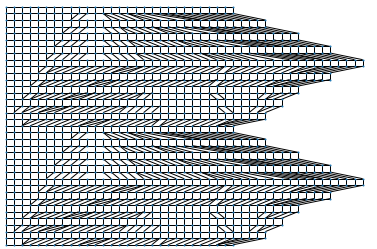}} \\ \hline
          \parbox[c]{\tabfig\textwidth}{Pincha \\ (one repeat) \\(960 nodes)} & \parbox[c]{0.12\textwidth}{ \includegraphics[width=\tabfig\textwidth,angle=90]{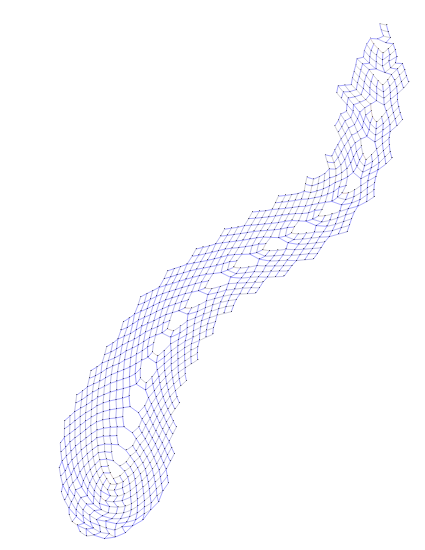}\\}  &\parbox[c]{\tabfig\textwidth}{\includegraphics[width=\tabfig\textwidth]{images/pincha.png}\\ }  
          & DNF
          & DNC \\ \hline
    \end{tabular}\medskip
    \caption{Overview of output from all algorithms on general knitting patterns.}
    \label{tab:edge_length_comparisons}
\end{table*}

% \begin{table}[]
%     \centering
%     \begin{tabular}{c|c|c}
%          Stitch & Row Size & Column Size  \\\hline\hline
%          k &  1 & 1.5 \\ 
%          yo & 0.6 & NA\\
%     \end{tabular}
%     \caption{Caption}
%     \label{tab:my_label}
% \end{table}

\begin{table*}[t]
    \centering
    \begin{tabular}{|c|c|c|c|c|}
         \hline Nodes  & DEL & SFDP & ImPrEd & KnitGrid\\ \hline \hline
         89 
         & \parbox[c]{\tabfig\textwidth}{\includegraphics[width=\tabfig\textwidth]{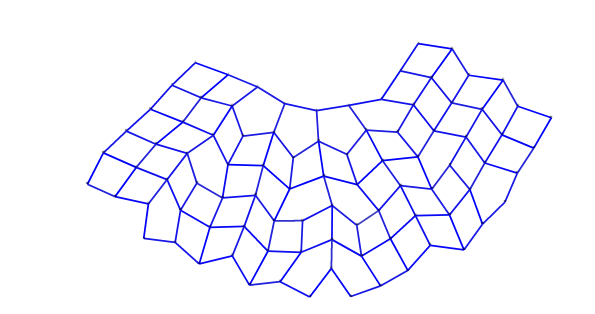} }  &\parbox[c]{\tabfig\textwidth}{\includegraphics[width=\tabfig\textwidth,angle=180]{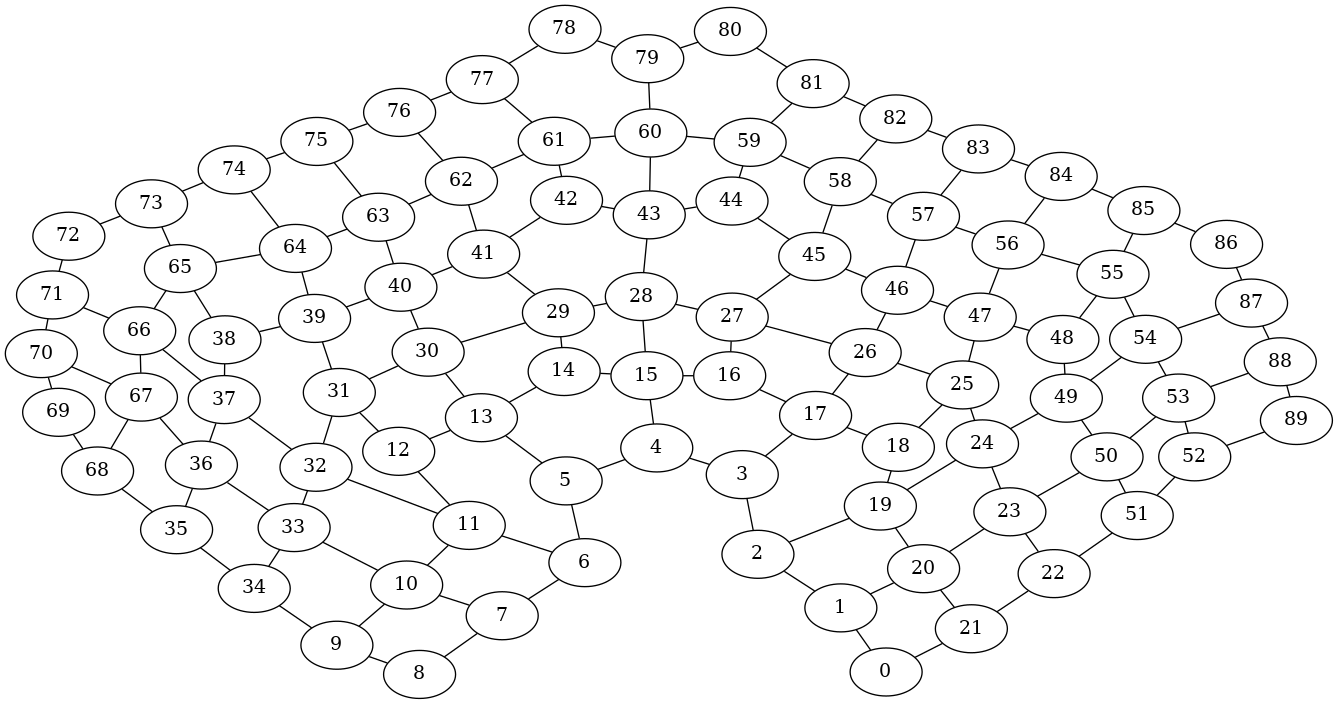}}  
          &\parbox[c]{\tabfig\textwidth}{\includegraphics[width=\tabfig\textwidth]{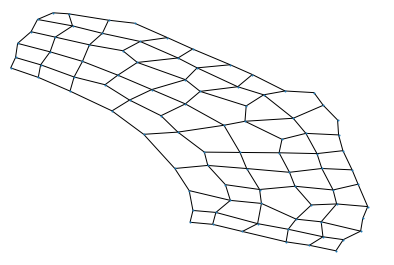}}
          &\parbox[c]{\tabfig\textwidth}{\includegraphics[width=\tabfig\textwidth]{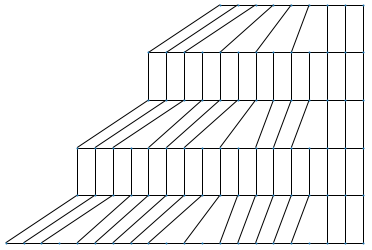}}\\ \hline
         285 
         &\parbox[c]{\tabfig\textwidth}{\includegraphics[width=\tabfig\textwidth]{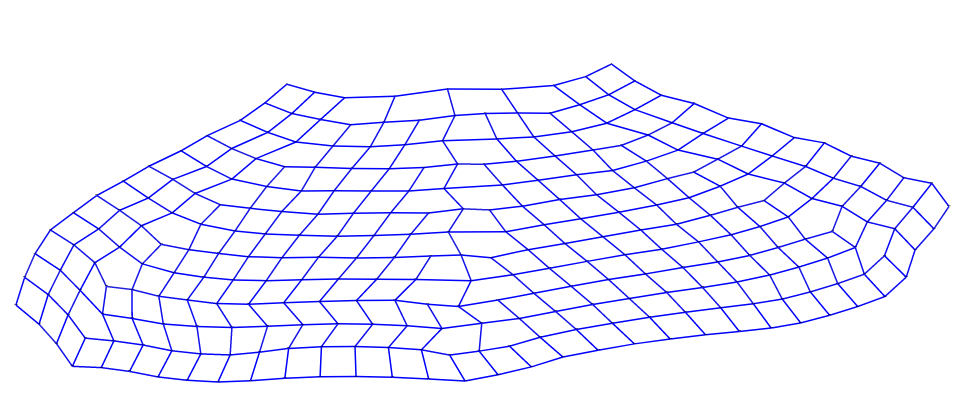}}
         &\parbox[c]{\tabfig\textwidth}{\includegraphics[width=\tabfig\textwidth]{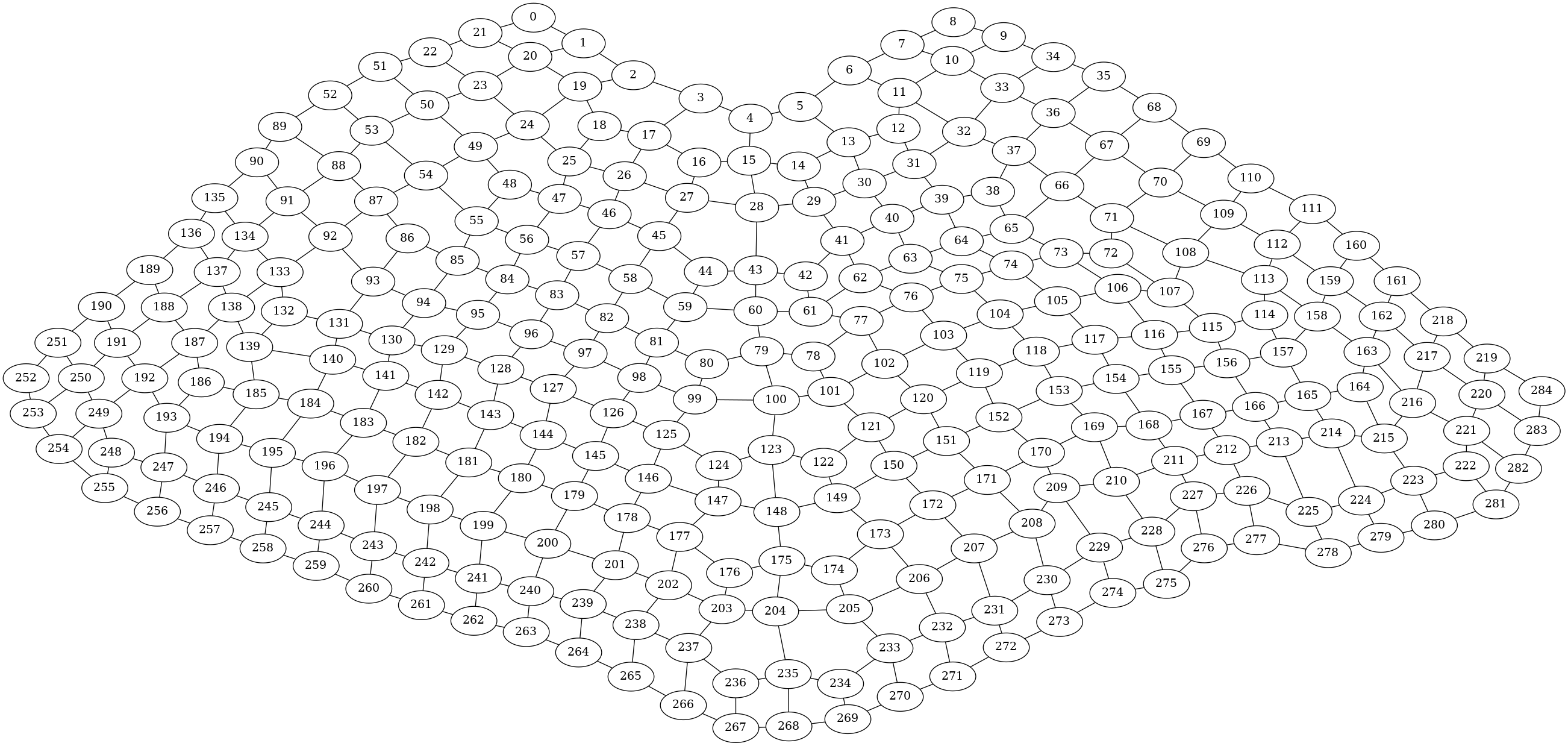}} &\parbox[c]{\tabfig\textwidth}{\includegraphics[width=\tabfig\textwidth]{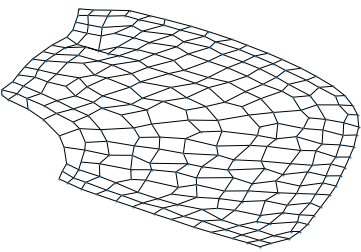}}
         &\parbox[c]{\tabfig\textwidth}{\includegraphics[width=\tabfig\textwidth]{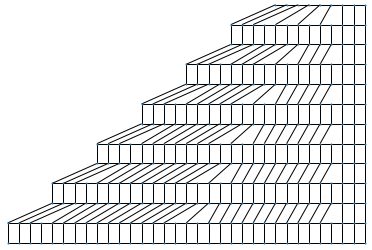}}\\ \hline
         513 
         & \parbox[c]{\tabfig\textwidth}{ \includegraphics[width=\tabfig\textwidth]{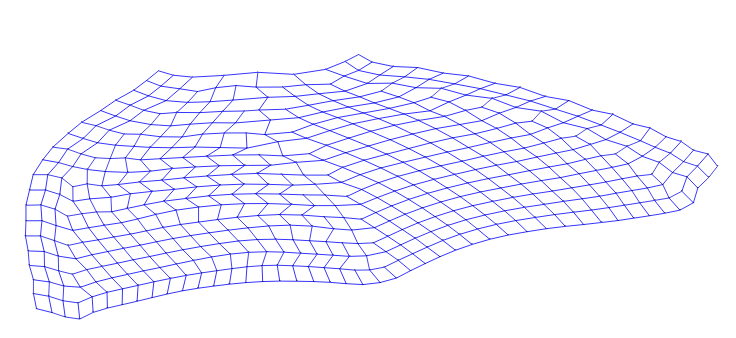}}&\parbox[c]{\tabfig\textwidth}{\includegraphics[width=\tabfig\textwidth,angle=180]{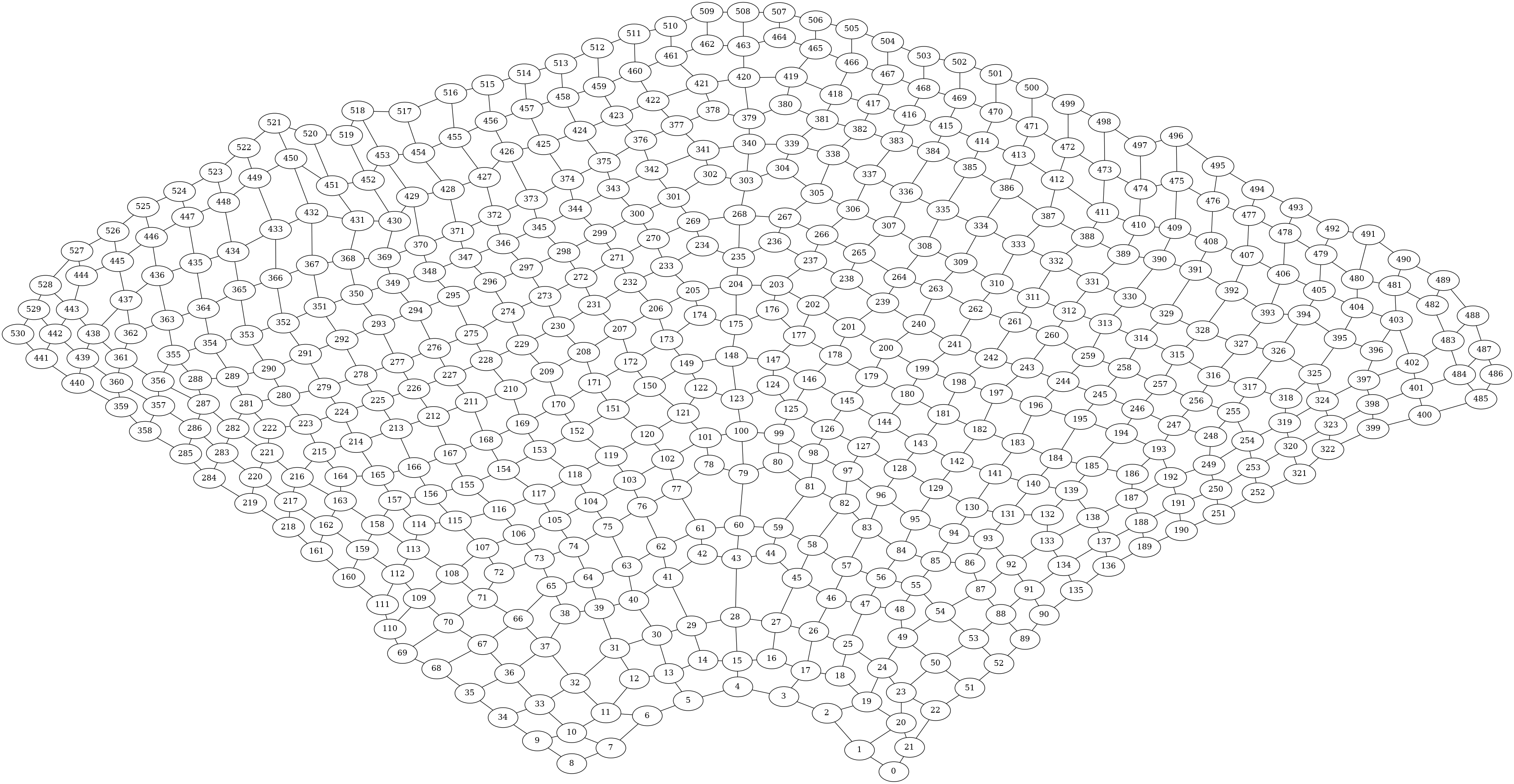}} 
         & \parbox[c]{\tabfig\textwidth}{DNF}
         & \parbox[c]{\tabfig\textwidth}{\includegraphics[width=\tabfig\textwidth]{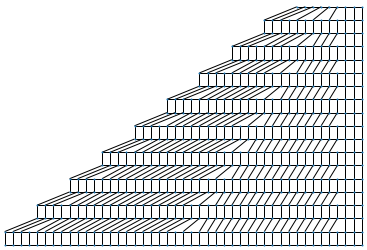}}\\ \hline
         845 & \parbox[c]{\tabfig\textwidth}{ \includegraphics[width=\tabfig\textwidth]{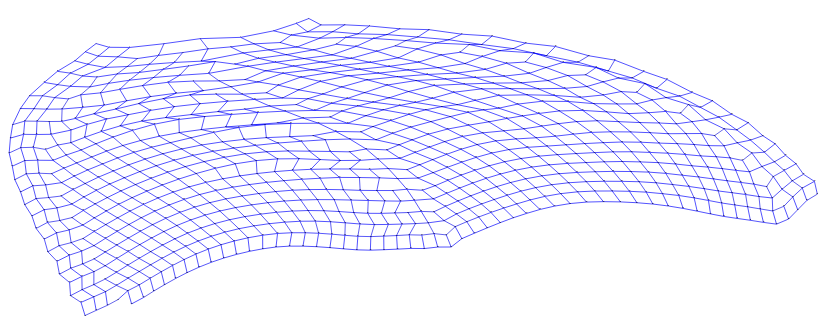}} 
         &\parbox[c]{\tabfig\textwidth}{\includegraphics[width=\tabfig\textwidth,angle=180]{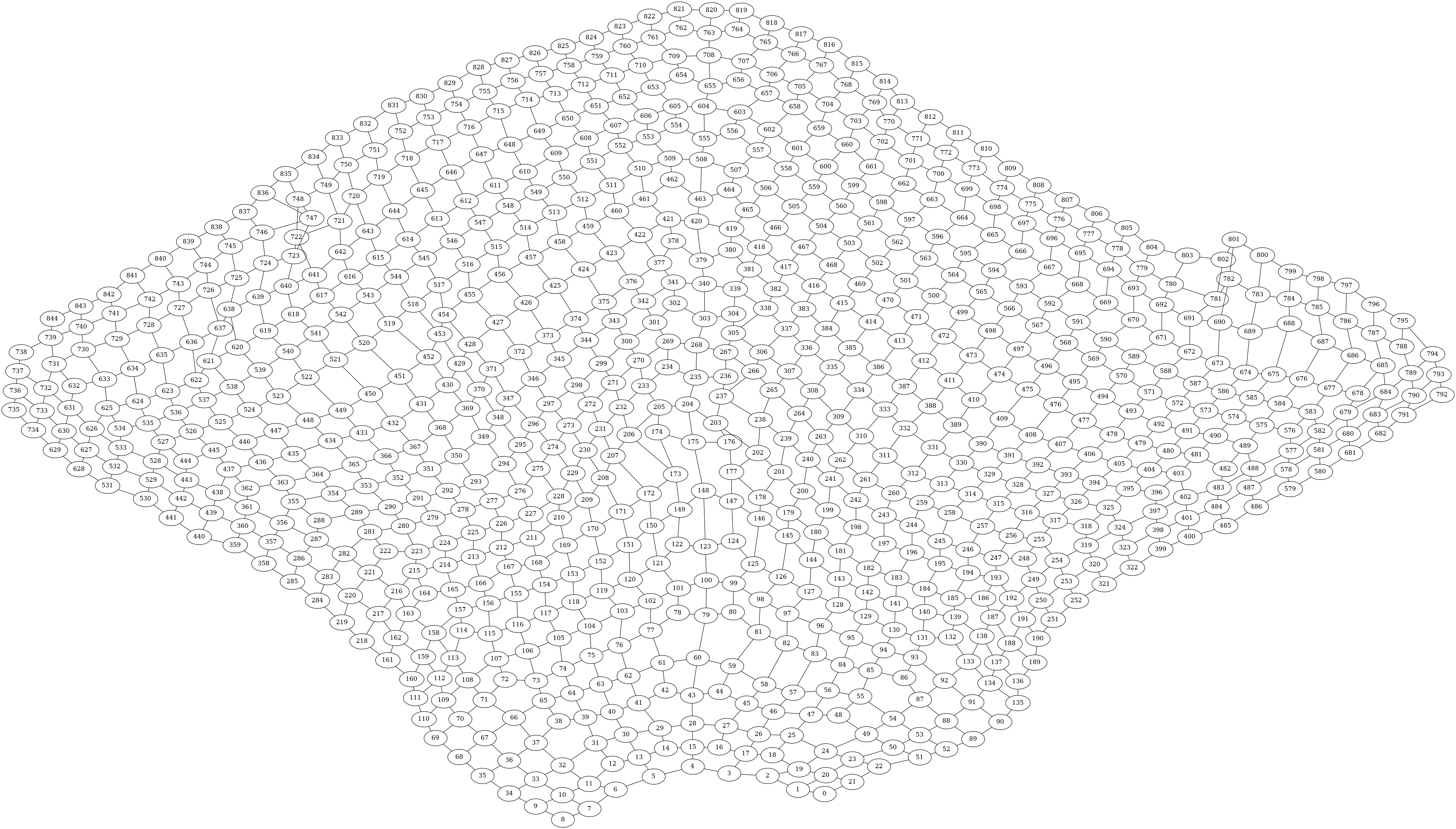}} 
         &\parbox[c]{\tabfig\textwidth}{DNF}
         &\parbox[c]{\tabfig\textwidth}{\includegraphics[width=\tabfig\textwidth]{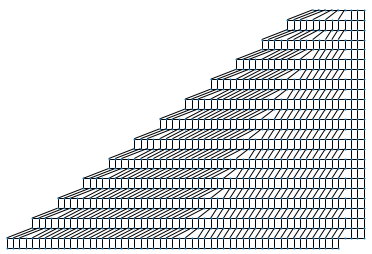}}\\ \hline
         1539 & \parbox[c]{\tabfig\textwidth}{ \includegraphics[width=\tabfig\textwidth]{images/triangle_sizes/triangle5.png}} 
         &\parbox[c]{\tabfig\textwidth}{\includegraphics[width=\tabfig\textwidth]{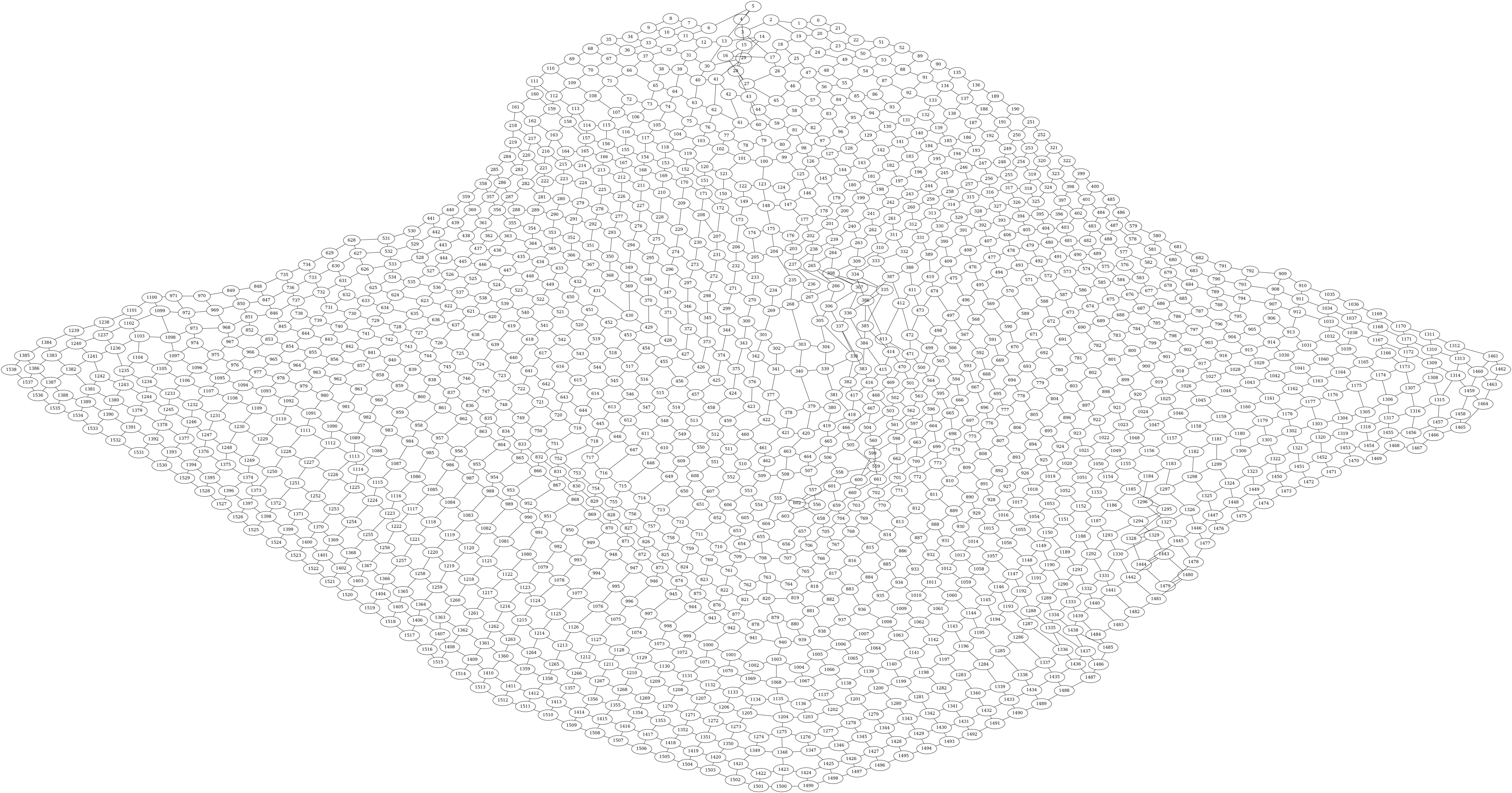}}
         & \parbox[c]{\tabfig\textwidth}{DNF}
         & \parbox[c]{\tabfig\textwidth}{\includegraphics[width=\tabfig\textwidth]{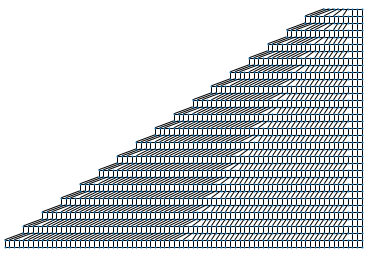}} \\ \hline
    \end{tabular}
    \medskip
    \caption{Triangle Graph -- Each row of the table includes six new knitting rows.}
    \label{tab:edge_length_comparisons}
\end{table*}

\section{Discussion and Limitations}

We propose a model for hand knitting patterns as graph representations. This  makes it possible to translate a huge variety of knitting patterns as  graph visualization tasks. We discuss classes of complexity in knit objects, all of which can be included in a parser with similar architecture. We focus only on visualizing our simplest complexity class which corresponds to planar knits and show that our force-directed approach outperforms prior methods.

One major limitation in this area is the lack of directly comparable algorithms. The need to preserve precise edge lengths, displaying the regular and complex structure of knitting techniques, and achieving crossing-free layouts leaves few viable tools. Additionally, we were unable to find a version of ImPrEd that produced crossing-free outputs when initialized with the planar layout. 

We hope that this paper will provide fertile ground for new graph layout algorithms. Knitting graphs provide  more structure than generic planar graphs, 1-planar graphs, or $k$-planar graphs. For example, the process of knitting imposes an ordering that should make it possible to create better initializations for the graph layout. This should help create better intial layouts by starting closer to the desired layout, although new layout restrictions may be required when extending to complexity classes 1 ($k-$planar graphs) and 2 (double-knitting graphs with multiple collocated vertices). %In both of these cases, edge lengths and stitch proximity will likely take precedence over number of crossings for representations. 

We hope that this paper and associated software (currently anonymized for the submission) represents a promising start in the direction of a useful tool for pattern-designers in the knitting community. A tool that converts knitting patterns into graphs and then visualizes the proposed patterns for a preview of the knitted output will provide for faster and easier examination of pattern shapes, reducing the amount of effort needed for testing patterns during the design process.

\section{Conclusions and Future Work}

In this paper, we present graph-drawing types of questions motivated by  knitting. In particular we provide a model to convert  hand-knitting patterns and into graphs with pre-scpedified edge lengths.  We also propose a layout algorithm that prevents edge crossings while optimizing edge lengths. 
This opens up a large fraction of hand-knitting patterns for automated representation and drafting assistance, however an overlapping large fraction of such patterns fall outside of the complexity class 0 (planar knitting) on for which our algorithms were built. 
%Furthermore, many patterns which do fit our complexity class 0 (planar knitting) are for garments and other shaped objects with significant curvature and shaping which make edge preserving planar representations infeasible. These patterns require 3 dimensional layouts for useful visualization. 

Natural follow up to this work would be  to generalize the layout algorithm proposed here to knitting patterns that go beyond the planar graphs considered here. %develop 3-d graph layouts on top of our existing representations. 
Another interesting question is thedevelopment of natural non-planar layouts which keep edge crossings in the location in which they would be naturally introduced during knitting. %Another natural next-step is the extension of evolving/dynamic layout algorithms (e.g. \cite{gray_evolving_2022}) to this problem, allowing users to add and remove stitches to a live layout, allowing rapid testing for small modifications. The rich structure of knitting makes node addition and subtraction easy to localize and thus a prime candidate for fast algorithms. An additional possibility is an algorithm focused on complexity classes 1 and 2 which admits hard and soft constraints e.g. based on UNICORN\cite{yu_unicorn_2022}. In this case the algorithm should be modified with numbers of crossings specified in the knitting pattern. 

%Knitting, an ancient fiber art, creates a structured fabric consisting of loops or stitches. Publishing hand knitting patterns involves lengthy testing periods and numerous knitters. Modeling knitting patterns with graphs can help expedite error detection and  pattern validation. In this paper, we describe how to model simple knitting patterns as planar graphs. We then design, implement, and evaluate a layout algorithm to visualize knitting patterns. Knitting patterns correspond to graphs with pre-specified edge lengths (e.g., uniform lengths, two lengths, etc.). This yields a natural graph layout optimization problem: realize a planar graph with pre-specified edge lengths, while ensuring there are no edge crossings. We quantitatively evaluate our algorithm using real knitting patterns of various sizes against three others; one created for knitting patterns, one that maintains planarity and optimizes edge lengths, and a popular force-directed algorithm. 

\newpage
%%
%% Bibliography
%%

%% Please use bibtex, 

\bibliography{bibliography}

\appendix

\end{document}